\documentclass[a4paper,11pt]{article}
\pdfoutput=1
\usepackage{jcappub}
\usepackage{journal_abbrev}
\usepackage{xparse,verbatim,color}
\usepackage{xspace,ulem,etoolbox,cprotect}
\xspaceaddexceptions{]\}}
\usepackage{bm}

\usepackage{graphicx}
\usepackage{adjustbox}
\usepackage{siunitx}
\usepackage{enumitem}
\usepackage{lmodern}

\usepackage{hyperref}

\def\fourmassbinplots#1{
  \centerline{\resizebox{\hsize}{!}{
      \includegraphics*{{plots/#1_lM12.5-13.0}.pdf}
      \includegraphics*{{plots/#1_lM13.0-13.5}.pdf}
      }}
  \centerline{\resizebox{\hsize}{!}{
      \includegraphics*{{plots/#1_lM13.5-14.0}.pdf}
      \includegraphics*{{plots/#1_lM14.0-14.5}.pdf}
  }}
}

\newcommand{\se}{\ensuremath{\sigma_\mathrm{8}}\xspace}
\newcommand{\AS}{\mathcal{A}_{s}}

\def\ba#1\ea{\begin{align}#1\end{align}}
\def\bea{\begin{eqnarray}}
\def\eea{\end{eqnarray}}
\def\be{\begin{equation}}
\def\ee{\end{equation}}
\def\d{\delta}

\def\({\left(}
\def\){\right)}
\def\[{\left[}
\def\]{\right]}
\def\<{\left\langle}
\def\>{\right\rangle}
\def\lapl{\nabla^2}
\def\vn{\boldsymbol{\nabla}}
\DeclareMathOperator{\tr}{tr}

\def\Ain{A_{\rm in}}
\def\hAin{\hat{A}_{\rm in}}

\newcommand{\perm}[1]{ \expandafter\ifstrempty\expandafter{#1} {\mbox{perm.}} {\mbox{$#1$ perm.}} }

\def\thH{\Theta_\text{H}}

\newcommand{\vs}{\nonumber\\}
\def\d{{\delta}}
\def\eps{{\varepsilon}}

\def\dgdet{\delta_{g,\rm det}}

\renewcommand{\v}[1]{\bm{#1}}

\def\vx{\v{x}}

\def\vk{\v{k}}
\def\vq{\v{q}}

\def\knl{k_\text{NL}}

\def\dirac#1{\delta_{\mathrm{D}}^{(#1)}}

\def\Mpch{\,h^{-1}\text{Mpc}}
\def\iMpch{\,h\,\text{Mpc}^{-1}}
\def\Msunh{\,h^{-1} M_\odot}
\def\Plin{P_\text{L}}
\def\nlin{n_{\text{L}}}
\def\P{\mathcal{P}}

\def\Lbox{L_\text{box}}
\def\L{\Lambda}
\def\Lin{\Lambda_\text{in}}

\def\W{\mathcal{W}}

\newcommand{\refeq}[1]{Eq.~(\ref{eq:#1})}

\newcommand{\reffig}[1]{Fig.~\ref{fig:#1}}

\newcommand{\reftab}[1]{Tab.~\ref{tab:#1}}
\newcommand{\refsec}[1]{Sec.~\ref{sec:#1}}

\newcommand{\refapp}[1]{Appendix~\ref{app:#1}}

\def\emph#1{\textit{#1}}
\def\bfem#1{\textit{#1}}

\title{Sigma-Eight at the Percent Level: The EFT Likelihood in Real Space}

\author[a]{Fabian Schmidt}

\emailAdd{fabians@mpa-garching.mpg.de}

\affiliation[a]{Max--Planck--Institut f\"ur Astrophysik, Karl--Schwarzschild--Stra\ss e 1, 85748 Garching, Germany}

\keywords{cosmological parameters from LSS, redshift surveys, dark matter halos, bias, effective field theory}

\arxivnumber{2009.14176}

\abstract{%
  The effective field theory likelihood for the density field of biased tracers allows for
  cosmology inference from the clustering of galaxies that consistently uses
  all available information at a given order in perturbation theory. This paper
  presents results and implementation details on the real-space
  (as opposed to Fourier-space) formulation
  of the likelihood, which allows for the incorporation of survey window
  functions. The implementation further uses a Lagrangian forward model
  for biased tracers which automatically accounts for all relevant contributions up to
  any desired order. Unbiased inference of $\sigma_8$ is demonstrated at
  the 2\% level for cutoff values $\L \leq 0.14 \iMpch$ for halo samples over
  a range of masses and redshifts. The inferred value shows the expected convergence to the ground truth in the low-cutoff limit. Apart from the possibility of including observational effects, this represents further substantial improvement over previous
  results based on the EFT likelihood.
}

\begin{document}

\maketitle

\flushbottom

\section{Introduction}
\label{sec:intro}

It is well established that the large-scale clustering of galaxies
contains a wealth of information on cosmology. This includes in particular
the angular diameter distance $D_A(z)$ and expansion rate $H(z)$ inferred from the baryon
acoustic oscillation (BAO) feature, the growth rate of structure
in form of the parameter combination $f \sigma_8$ inferred from redshift-space
distortions (RSD), and the amplitude of local-type primordial non-Gaussianity $f_{\rm NL}$ 
via the large-scale scale-dependent bias. All of these constraints
are accessible using \emph{linear information}, i.e. using linear theory
model predictions for the galaxy power spectrum or 2-point correlation
function, although nonlinear corrections are important to take into account
essentially as theoretical error bar \cite{Baldauf:2016sjb,Nishimichi:2020tvu,Chudaykin:2020hbf}.

However, there is significantly more information to be exploited in the
nonlinear part of galaxy clustering. An example is BAO reconstruction,
which reduces the inferred error bars on $D_A, H$ by performing a nonlinear
operation on the data (e.g., \cite{eisenstein/seo/etal:2007,2009PhRvD..79f3523P}).
Another is the normalization of the
linear matter power spectrum \se, which at linear order is perfectly
degenerate with the linear galaxy bias parameter $b_1$
(see Secs.~1-2 of \cite{biasreview} for an introduction to the topic of bias). Including nonlinear
information, for example via the bispectrum, allows for this degeneracy
to be broken. Combined with RSD, this in turn allows for independent
constraints on the growth rate proper $f = d\ln D/d\ln a$ where $D(a)$
is the linear growth factor, and \se, allowing for interesting additional
tests of the $\L$CDM paradigm and possibly improved constraints on the sum of neutrino masses. It is also
worth noting that selection effects can spoil the determination of $f \sigma_8$
at linear order \cite{Zheng:2010jf,Wyithe:2011mt,Behrens:2017xmm,Krause:2010tt,Martens:2018uqj,Obuljen:2019ukz,Desjacques:2018pfv}, in which case nonlinear information is likewise
necessary in order to extract the growth rate \cite{Agarwal:2020lov}. 

Linear information is by definition optimally extracted via the
power spectrum. A new approach to extracting the nonlinear information,
which is only marginally accessed by the power spectrum, has recently gathered increased attention: Loosely referred to as
\emph{forward modeling}, this approach proceeds by writing down a likelihood for the entire galaxy density field,
and performing a full joint Bayesian inference of cosmological parameters
along with the phases of the initial conditions corresponding to the
observed survey volume
\cite{1989ApJ...336L...5B, 1994ApJ...423L..93L, 1995MNRAS.272..885F,
	1999AJ....118.1146S, 2004MNRAS.352..939E, 2010MNRAS.406...60J,
	2010MNRAS.407...29J, 2010MNRAS.409..355J, 2010MNRAS.403..589K,
	2012MNRAS.420...61K, 2013MNRAS.432..894J, Wang:2014hia, 2015MNRAS.446.4250A,
	Wang:2016qbz, 2017MNRAS.467.3993A,2017JCAP...12..009S,
        schmittfull/etal:2018,Modi:2019hnu}.
The crucial ingredient in this inference approach is the conditional probability (``likelihood'')
of the observed galaxy density field given the forward-evolved matter
density field. Unbiased inference depends very sensitively on the accuracy of
the likelihood for the galaxy density field.
Refs.~\cite{paperI,paperII,cabass/schmidt:2019} recently derived the likelihood
in the context
of the effective field theory (EFT \cite{baumann/etal:2012,carrasco/etal:2012}) of large-scale structure. By including only modes with wavenumbers below
a cutoff, or maximum wavenumber $\L$, this allows for a rigorous, controlled
Bayesian inference of cosmology and initial conditions, provided that
the cutoff is chosen to only include modes for which perturbation theory
is valid ($\L \lesssim 0.25\iMpch$ at $z=0$).

Ref.~\cite{paperIIb} presented results of the EFT likelihood applied
to halo samples in their rest frame (i.e. without RSD) identified in N-body
simulations. In particular, unbiased inference of \se (equivalent, when all remaining cosmological parameters are fixed, to the primordial normalization $\sqrt{\mathcal{A}_s}$)
was demonstrated
at the 4--8\% level depending on halo mass and redshift. In this test, the
phases of the forward model were fixed to the truth, i.e. the initial conditions used
in the N-body simulations. As mentioned above,
the constraint on \se is not obtainable using linear theory due to the bias-amplitude
degeneracy, and so is exclusively due to the robust extraction of nonlinear
information. These results were based on a Fourier-space formulation of
the EFT likelihood.

In this paper, we present an implementation of the EFT likelihood in real space based on \cite{cabass/schmidt:2020}.
As shown in Ref.~\cite{cabass/schmidt:2020}, this formulation allows for an
incorporation of the most important corrections to the leading EFT likelihood, a Gaussian with constant diagonal noise, in particular the modulation of the noise by long-wavelength modes, or equivalently the stochasticity of the bias parameters. Moreover, this formulation also permits one to include the survey window function in a computationally efficient way. We describe how this works in \refsec{like}.

We further employ a new Lagrangian forward model to predict the mean galaxy density given fixed phases. This forward model is recursively constructed to any desired order based on Lagrangian perturbation theory (LPT), without assuming
the Einstein-de Sitter approximation that is usually adopted. It also
includes all or a large fraction of the relevant higher-derivative contributions
(depending on order).
Apart from the efficient, recursive construction to any order,
the Lagrangian forward model has the further advantage of a straightforward
transformation to redshift space and the lightcone, i.e. onto the observer's past
lightcone rather than a constant proper time slice. 
We briefly describe the forward model in \refsec{lptbias}, relegating a more detailed exposition to the upcoming Ref.~\cite{paper_nLPT}. 

To summarize, \bfem{this paper presents an implementation of the EFT likelihood
that allows for the efficient incorporation of perturbation theory
contributions up to any desired order,\footnote{Given computational constraints, in particular memory requirements.} as well as the leading observational
effects, namely RSD, lightcone, and window function.}

In our results, we will again restrict to the \se inference from rest-frame
halo catalogs, deferring the inclusion of RSD, lightcone and window function to upcoming
work. Thus, our goal will be to test the convergence of perturbation theory
as a function of the cutoff $\L$ at the level of the \se inference,
and the extent to which going to higher order improves the latter.
\reffig{alpha_vs_X} shows a summary of the results, namely the residual fractional error in $\se$, $\hAin-1$, for all halo samples and redshifts at a fixed cutoff $\L$ and for different expansion orders. 

The paper is organized as follows. In \refsec{like} we present the
real-space EFT likelihood, including subtleties such as the implementation
of the sharp-$k$ cut, and discuss the incorporation of the window function.
\refsec{lptbias} briefly presents the forward model and bias expansion,
including the ordering scheme used to determine the relevant higher-derivative
and stochastic operators. \refsec{impl} presents details on the numerical
implementation, including marginalization of the (numerous) bias parameters.
We then turn to the results in \refsec{results}, and their discussion in
\refsec{disc}. We conclude in \refsec{conc}. The appendices provide additional
figures, and more
details on the halo samples and numerical implementation.

\section{The EFT likelihood in real space}
\label{sec:like}

The goal of the forward-modeling approach to galaxy clustering is to evaluate
the posterior $\P(\theta|\d_g)$ of the cosmological parameters $\theta$ given the
observed galaxy density \emph{field} $\d_g(\vx)$. This is obtained
by marginalizing the likelihood $\P(\d_g | \theta, \{b_O\}, \d_{\rm in})$
over the phases $\d_{\rm in}$ corresponding to the primordial fluctuations
in the volume covered by the galaxy survey, which are drawn from a multivariate Gaussian prior $\P(\d_{\rm in}|\theta)$, as well as nuisance parameters
such as the bias coefficients $\{ b_O\}$. The key physical ingredient in this
approach thus is the likelihood $\P(\d_g | \theta, \{b_O\}, \d_{\rm in})$.
The contribution of \cite{paperI,paperII,cabass/schmidt:2019,paperIIb} was to
derive this likelihood in the context of the effective field theory of LSS.

Before turning to the likelihood, let us consider the specific question
we will investigate numerically in this paper. The highly nontrivial and
numerically costly marginalization over the phases can be avoided by
looking at simulations: then we can fix $\d_{\rm in}$ to the known initial
density field of the simulations. Further, studying physical tracers
(in this case, halos) in their rest frame allows us to drop redshift-space
distortions from the forward model. Finally, we restrict the set of
cosmological parameters $\theta$ to the linear power spectrum normalization
\se; equivalently, the square-root of the primordial amplitude $\sqrt{\mathcal{A}_s}$.
Thus, the question we investigate below,
the same question as considered in \cite{paperI,paperII,paperIIb},
is: \bfem{How well can one
  recover $\se$ from a rest-frame tracer catalog, with no prior knowledge on the selection, i.e. bias parameters, but
  assuming perfect knowledge of the phases of the initial conditions?}
This question is highly nontrivial, since \se is perfectly degenerate
with the linear bias $b_1$ in linear theory (i.e. on large scales). Thus,
all constraints obtained in \se are  \bfem{based purely on nonlinear information.}

Let us now turn to the likelihood $\P(\d_g | \se, \{b_O\}, \d_{\rm in})$,
simply referred to as ``likelihood'' in the following,
which we need for this study. 
As discussed at length in \cite{paperI,paperII,cabass/schmidt:2019,paperIIb},
the likelihood for the density field of a biased tracer $\d_g(\vx)$
consists of
\begin{enumerate}
  \item[$(i)$] a model for the nonlinear evolution of the matter density
    under gravity, which we denote as $\d_{\rm fwd}[\d_{\rm in}]$;
  \item[$(ii)$] a deterministic bias expansion of the biased tracer, which we denote as
    $\dgdet[\d_{\rm in}]$, and which incorporates ingredient $(i)$; and
  \item[$(iii)$] the log-likelihood $\ln\P(\d_g | \dgdet)$ for the observed galaxy
    field, i.e. the data $\d_g(\vx)$ given the deterministic prediction $\dgdet$.
\end{enumerate}
Note that $\dgdet$ depends on a set of bias parameters $b_O$ which
in general have to be determined from the data. 
In this section we focus on ingredient $(iii)$, and turn to the bias
expansion employed here in \refsec{lptbias}.

Refs.~\cite{paperI,paperII,cabass/schmidt:2019,paperIIb} were based
on an expression for the likelihood given in Fourier space,
\ba
\ln\P(\d_g | \dgdet) = -\frac12 \int_{|\vk|<\Lambda} \frac{d^3\vk}{(2\pi)^3}
\left\{\left[\frac{|\d_g(\vk) - \dgdet(\vk)|^2}{P_\eps(k)}
  + \ln [2\pi P_\eps(k)]\right]\right\}
\label{eq:lnPF}
\ea
where the noise variance is parametrized as
\ba
P_\eps(k) = P_\eps^{\{0\}} + P_\eps^{\{2\}} k^2 + \ldots ,
\label{eq:Peps}
\ea
and $\L$ is the cutoff in wavenumber. 
If we set the subleading contribution $P_\eps^{\{2\}}$, as well as all
higher-order terms in the expansion \refeq{Peps}, to zero, then \refeq{lnPF}
can be formally transformed to a real-space likelihood, by way of Parseval's theorem:
\ba
\ln\P(\d_g | \dgdet) \stackrel{P_\eps={\rm const}}{=} -\frac12 \int d^3 \vx
\left[\frac{\left(\d_{g,\L}(\vx) - \d_{g,\rm det,\L}(\vx)\right)^2}{P_\eps^{\{0\}}} \right]
-\frac12 N_{\rm modes} \ln [2\pi P_\eps^{\{0\}}]
\label{eq:lnPR}
\ea
where the
fields $\d_{g,\L}$ and $\d_{g,\rm det,\L}$ are sharp-$k$ filtered at the cutoff $\L$,
i.e. 
\be
\d_{g,\L}(\vk) \equiv \d_g(\vk) W_\L(\vk)
\label{eq:dgL}
\ee
and similarly for $\d_{g,\rm det,\L}$. The filter choice corresponding to
\refeq{lnPF} is $W_\L(\vk) = \thH(\L - |\vk|)$, where $\thH$ is the Heaviside function. 
$N_{\rm modes}$ is the number of modes that survive the sharp-$k$ cut
(hence this number is proportional to the volume),
ensuring the equivalence to \refeq{lnPF}.\footnote{When discretized on a grid, one should also replace $P_\eps^{\{0\}} \to N_g^3 P_\eps^{\{0\}}$, where $N_g$ is the number of grid points in one dimension.}

A real-space expression for the likelihood offers several advantages.
First, as argued in \cite{cabass/schmidt:2020}, the leading perturbative
corrections to the likelihood beyond the constant-noise expression \refeq{lnPR}
are in fact the
modulation of the noise amplitude by long-wavelength perturbations
(field-dependent stochasticity)
rather than the higher-derivative terms in \refeq{Peps}; schematically,
these are given in real space by
\be
P_\eps(\vx) = P_\eps^{\{0\}} + P_{\eps,\eps_\d} \d(\vx) + \ldots .
\label{eq:PepsR}
\ee
Physically, it is expected that the noise in the tracer field is different
in high- vs. low-density regions. This effect is captured by \refeq{PepsR}.
An equivalent interpretation is that it encodes stochasticity in the bias
parameters; specifically, the term in \refeq{PepsR} corresponds to
stochasticity in the linear bias $b_1 = b_\delta$. 
  The reason that the term $\propto \d(\vx)$ in \refeq{PepsR} is more relevant than the $k^2$ term in \refeq{Peps} is the shape of the matter power spectrum, since $\d(\vx) \sim [k^3 \Plin(k)/2\pi^2]^{1/2}_{k=\Lambda} \propto \Lambda^{(3+\nlin)/2}$, if the linear matter power spectrum $\Plin(k) \propto k^{\nlin}$, while the higher-derivative term in \refeq{Peps} scales as $\Lambda^2$. For $\nlin\approx -1.5$, the former is more relevant.

The second significant advantage of a real-space likelihood is that it
allows for a more direct incorporation of the window function of an actual
galaxy survey. Consider the simplest case of a binary window function,
where $\W(\vx)=1$ if the region containing $\vx$ is observed, and $\W(\vx)=0$ otherwise (this
includes the case where we do not trust the selection of galaxies in a given
region e.g. due to flux calibration issues caused by bright nearby stars).
In the context of a Bayesian forward model, $\W(\vx)=0$ corresponds to setting
$P_\eps(\vx)$ to infinity; thus, we can generalize \refeq{lnPR} to include the window function:
\ba
\ln\P(\d_g | \dgdet) = -\frac12 \int_{\W(\vx) > 0} d^3 \vx
\left[\frac{\left(\d_g^\L(\vx) - \dgdet^\L(\vx)\right)^2}{P_\eps^{\{0\}}} \right]
+ {\cal N} \quad
\mbox{(binary window function $\W$)},
\ea
where ${\cal N}$ is a normalization constant. 
Notice that, unlike the expansions in \refeq{Peps} and \refeq{PepsR},
the window function is a \emph{non-perturbative} effect. We defer
an actual implementation of window functions to future work, restricting
this paper to simulations with trivial window functions $\W=1$. 
To summarize, we expect that a real-space likelihood will capture both
the \emph{dominant observational} effect as well as the \emph{leading perturbative} corrections to the EFT likelihood.

Unfortunately, the actual implementation is not quite as simple as
\refeq{lnPR}. In order to obtain a practical real-space likelihood, we
need to satisfy two requirements: a field-level covariance that is diagonal
in real space (otherwise, the likelihood would not simply be given by a single
integral over real space as in \refeq{lnPR}); and a sharp-$k$ filter which
ensures that only modes with $|\vk|<\L$ appear in the likelihood.
We can think of the second requirement in Fourier space, again specializing to a
constant noise covariance $P_\eps(k) = P_\eps^{\{0\}}$, as a diagonal
covariance with a step-function behavior:
\be
   {\rm Cov}(\vk,\vk') = (2\pi)^3 \dirac3(\vk+\vk') P_\eps(k)\Big|_\L
   \quad\mbox{where}\quad
P_\eps(k)\Big|_\L = \left\{
\begin{array}{ll}
  P_\eps^{\{0\}}, & k < \L \\
  \infty, & k \geq \L
\end{array}
\right.
.
\ee
If we transform this covariance into real space, the result is not diagonal \cite{cabass/schmidt:2020}. This is in keeping
with Parseval's theorem, since we cannot invoke it to go from \refeq{lnPF}
to \refeq{lnPR} if $P_\eps$ is not constant.

An alternative opens up if we move away from the spherical sharp-$k$ cut to a
cubic cut,
\be
W_\L(\vk) = \thH(\L - |\vk|) \longrightarrow \prod_{i=1}^3 \thH(\L - k_i),
\label{eq:WL}
\ee
where $k_i$ denote the Cartesian components of the Fourier vector $\vk$.
Notice that the maximum wavenumber allowed by the cubic filter is
$\sqrt{3} \L$. 
In this case, we can use a  discrete Fourier representation of the
filtered fields (\refeq{dgL}) 
which preserves \emph{precisely} those modes that are nonzero after
the cubic sharp-$k$ filter in \refeq{WL}. This is achieved by choosing
a grid size $N_g$ such that $k_{\rm Ny}(N_g)/2 = \L$, where $k_{\rm Ny}$
is the Nyquist frequency of the grid; that is,
\be
N_g(\L) = \left\lfloor\frac{\Lbox \L}{\pi}\right\rfloor,
\label{eq:NgL}
\ee
where $\Lbox$ is the side length of the cubic box in real space. More precisely,
we choose the largest even number that is smaller than $\Lbox\L/\pi$.
In the case studied here, $\Lbox^3$ corresponds to the simulation volume, while in an
application to real data this would be the reconstruction volume which
encompasses the entire survey. Note that \refeq{NgL} restricts the actual
cutoff to discrete values; however, if $\Lbox \L \gg 1$, as is the
case in practical applications, this is a minor restriction.

The filter \refeq{WL} breaks rotational invariance, i.e. it introduces preferred
directions. For simulated objects, this is not expected to be an issue,
since there are no intrinsic directions in the simulations that the coordinate axes could align with.
In the application to real data, it is possible that the alignment of the coordinate
axes with preferred directions in the survey volume could lead to small systematic
artefacts. This can however be tested for by rotating the coordinate axes of the
reconstruction volume.

Once the fields are discretized on a Fourier-space grid of size $N_g(\L)$,
one can then transform them back to real space to obtain representations, e.g. 
$\d_{g,\L}(\vx)$, which only contain modes below the cutoff.
It is then appropriate to use a diagonal, real-space likelihood in
terms of $\d_{g,\L}(\vx), \d_{g,\rm det,\L}(\vx)$ which can incorporate
the field-dependent stochasticity in \refeq{PepsR}, as well as the window function.
The details of the implementation are described in \refsec{impl}.
In this paper, we will restrict to a constant covariance in real space,
for reasons discussed in that section.

\section{Construction of biased field}
\label{sec:lptbias}

The deterministic mean-field prediction for the galaxy density can generally be
written as
\be
\dgdet(\vx,\tau) = \sum_O b_O(\tau) O(\vx,\tau),
\ee
where $b_O$ denote the bias coefficients, and the operators $O$ are in general
constructed from the second-derivative tensor of the gravitational potential,
$\partial_i\partial_j\Phi$, and spatial derivatives thereof.
The operators are designed to span the entire set of local gravitational
observables, and are ordered in terms of perturbations (powers of $\Phi$) and
number of spatial derivatives, so that there is only a finite number of operators
relevant at a given order (we will return to the precise ordering in \refsec{hderiv}).
The goal then is to obtain constraints on cosmological parameters after
marginalizing over the bias parameters $b_O$, which in general are unknown for
a given observed LSS tracer (certainly not known to the percent level required
for precision inference of cosmological parameters). The same set of operators also appears in the
expansion of the real-space covariance, \refeq{PepsR}.

In this section, we describe the construction of the operators $O(\vx)$.
The previous papers in this series \cite{paperI,paperII,paperIIb} used a \emph{Eulerian} bias expansion (but, unlike e.g. \cite{gridSPT}, built
on the LPT matter density). That is, the bias operators were constructed out of the forward-evolved matter density field. Here, we instead use a Lagrangian bias expansion, which first constructs the bias operators and then displaces them to the Eulerian frame. This is very similar to the approach recently described in \cite{schmittfull/etal:2018}, and has several advantages as mentioned in \refsec{intro}. We provide an outline here, with more details being presented in the upcoming Ref.~\cite{paper_nLPT}.

\subsection{Lagrangian bias expansion}

The Lagrangian density of any biased tracer can be expanded as
\be
\dgdet^{\rm L}(\vq,\tau) = \sum_O b_O^{\rm L} O^{\rm L}(\vq,\tau),
\label{eq:dgdetL}
\ee
where a superscript L indicates quantities in the Lagrangian (fluid rest) frame.
More precisely, $\vq$ denotes the initial positions of matter particles as $\tau\to 0$.
The relation to the final observed position is given by the Lagrangian displacement
$\v{s}$, via
\be
\vx(\tau) = \vq + \v{s}(\vq,\tau).
\label{eq:vs}
\ee
Since the density perturbations at the initial time are vanishingly small, the
late-time matter density field is completely described by the displacement field $\v{s}(\vq,\tau)$, which can be expanded in orders of perturbations:
\be
\v{s}(\vq,\tau) = \v{s}^{(1)}(\vq,\tau) + \v{s}^{(2)}(\vq,\tau) + \cdots .
\ee
The equations of motion of $\v{s}$ can be integrated to give convenient recursion
relations that allow for a relatively simple numerical computation of the $\v{s}^{(n)}$ \cite{rampf:2012,zheligovsky/frisch,matsubara:2015,paper_nLPT}.

As argued in \cite{MSZ} (see also Sec.~2.5 of \cite{biasreview}), 
the set of Lagrangian bias operators $O^{\rm L}$,
at leading order in spatial derivatives, comprises all scalar combinations
 of the contributions $\v{M}^{(n)}$ to the symmetric part of the Lagrangian distortion tensor
 \be
M_{ij}(\vq,\tau) \equiv \partial_{q,(i} s_{j)}(\vq,\tau),
\ee
with the exception of $\tr[ \v{M}^{(n)} ]$ with $n>1$, which is degenerate with the other terms. 
 Since the antisymmetric (transverse) part of the distortion tensor does not
 appear in the bias expansion, we only consider the symmetric part in this paper (see \cite{paper_nLPT} for details).

 Using the fact that any symmetric $3\times 3$ matrix only contains 3 linearly independent rotational invariants, the complete Lagrangian basis up to fourth order is then given by \cite{biasreview}
 \ba
 O^{\rm L} \in
 \left\{
 \begin{array}{ll}
\left({\rm 1^{st}}\right) \ & \ \tr[M^{(1)}]   \\[4pt] 
\left({\rm 2^{nd}}\right) \ & \ \tr[(M^{(1)})^2]\,,\  (\tr[M^{(1)}])^2 \\[4pt] 
\left({\rm 3^{rd}}\right) \ & \ \tr[(M^{(1)})^3 ]\,,\ \tr[(M^{(1)})^2 ]  \tr[M^{(1)}],\ (\tr[M^{(1)}])^3\,,\ \tr[M^{(1)} M^{(2)}] \\[4pt] 
\left({\rm 4^{th}}\right) \ & \ \tr[(M^{(1)})^3 ]  \tr[M^{(1)}]\,,\
\tr[(M^{(1)})^2 ](\tr[M^{(1)} ])^2\,,\
\left(\tr[(M^{(1)})^2 ]\right)^2\,,\ (\tr[M^{(1)}])^4\,,\\
& \ \tr[M^{(1)}] \tr[M^{(1)} M^{(2)}]\,,\  \tr[M^{(1)} M^{(1)} M^{(2)}]\,,
\ \tr[M^{(1)} M^{(3)}]\,, \   \tr[M^{(2)} M^{(2)}] \;.
\end{array}
\right. 
\label{eq:LagrBasis}  
\ea
Moreover, the construction can be straightforwardly continued to higher order. The code implementation in fact provides a construction to any desired order.

\refeq{LagrBasis} is only complete at fourth order when assuming the EdS approximation. In a general background, there is one additional term; specifically $\tr [\v{M}^{(3)} \v{M}^{(1)}]$ generalizes to
\be
\tr [\v{M}^{(3,1)} \v{M}^{(1)}], \quad \tr [\v{M}^{(3,2)} \v{M}^{(1)}],
\ee
where $\v{M}^{(3,i)}$ denote two different shapes corresponding to distinct time evolution at third order (see App.~C of \cite{biasreview} for the corresponding terms in the Eulerian bias expansion).
We have found that the splitting of $\tr [\v{M}^{(3)} \v{M}^{(1)}]$ has an entirely negligible numerical impact in our analysis, which is why we restrict to the EdS expansion \refeq{LagrBasis} for the results in this paper. Again, the code implementation allows for the fully general bias construction for any expansion history \cite{paper_nLPT}.
For comparison, we will however show results that employ $n$-th order LPT beyond the EdS approximation, where we insert the total $\v{M}^{(3)}$ into the bias expansion \refeq{LagrBasis} (even in that case, we find extremely small differences to the EdS approximation).

Based on Lagrangian recursion relations \cite{rampf:2012,zheligovsky/frisch,matsubara:2015}, the tensors $\v{M}^{(n)}$ are constructed recursively starting from
\be
M^{(1)}_{ij} = \frac{\partial_q^i\partial_q^j}{\lapl_q} (\vn\cdot\v{s}^{(1)})
= - \frac{\partial_q^i\partial_q^j}{\lapl_q} \delta^{(1)},
\ee
where $\delta^{(1)}$ is the linear density field.
As discussed in \cite{paperIIb}, a sharp-$k$ cutoff $\Lin$ is to be imposed in the
initial density field. This should be greater or equal to the cutoff $\L$ in the
likelihood, since the likelihood will only be valid if at least all linear modes
are represented. We have not found any significant improvement when choosing
$\Lin > \L$, so we adopt the most well motivated choice of $\Lin=\L$, 
\be
\delta^{(1)}(\vk) \to \delta^{(1)}_\L(\vk) = W_\L(\vk) \delta^{(1)}_\L(\vk),
\ee
where here the filter is the cubic sharp-$k$ filter given in \refeq{WL}. 
Given the set of $\{\v{M}^{(i)}\}_{i=1}^{n-1}$, the bias expansion up to order
$n$ is then constructed by taking all products of all invariants of
the $\v{M}^{(i)}$ up to order $n$ (equivalently, up to $n=4$, all terms in \refeq{LagrBasis}). This yields the Lagrangian-space deterministic galaxy density field,
\refeq{dgdetL}. 

\begin{figure*}[t]%
  \includegraphics[width=\textwidth]{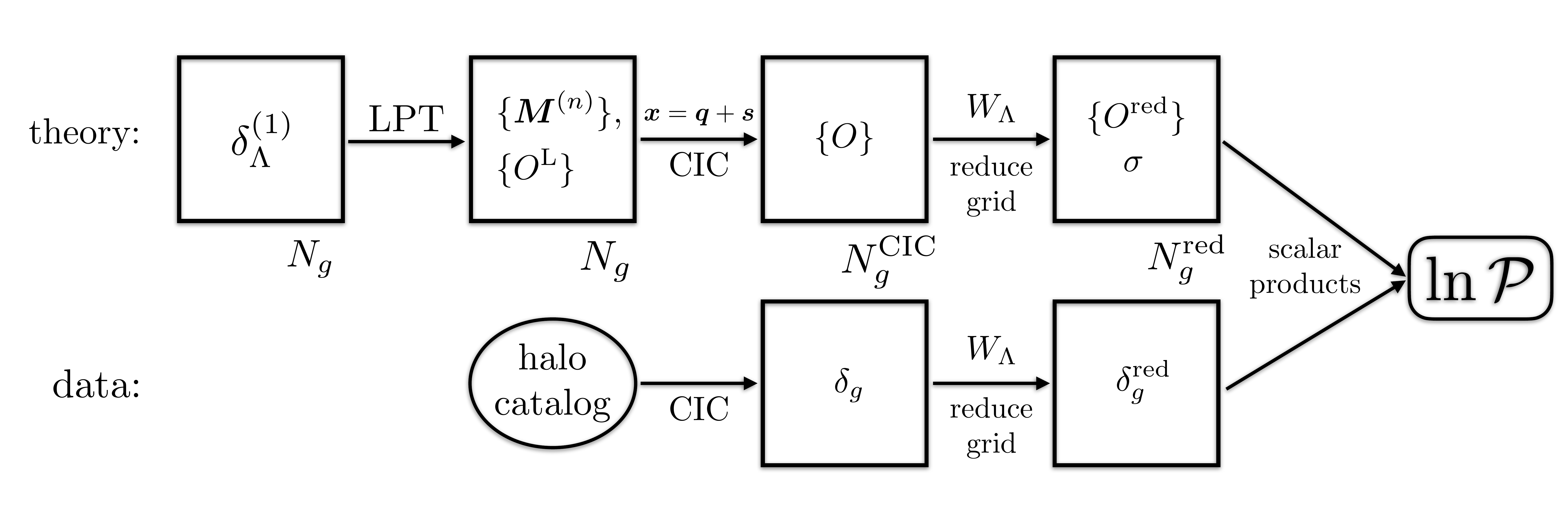}
  \cprotect\caption{Flowchart of the steps involved in the computation of the forward model and likelihood. Three different grids are involved: $N_g$ is chosen to have a Nyquist frequency equal to or larger than $n \Lambda$, where $n$ is the order in perturbations of the forward model or bias expansion, whichever is larger; $N_g^{\rm CIC}=512$ is fixed; and $N_g^{\rm red}$ has a Nyquist frequency equal to $\L$ (\refeq{NgL}). Both real and Fourier representations are involved on each grid.
    \label{fig:flowchart}}
\end{figure*}

We now need to transform, or displace, the galaxy density field to Eulerian coordinates
via \refeq{vs}. In order to be able to marginalize over the bias parameters in the
end, we in fact displace the Lagrangian bias operators individually. This proceeds
as follows. Each operator (except the linear-order $\tr [\v{M}^{(1)}]$, see below),
is copied to a larger grid of size $(N_g^{\rm CIC})^3$ in Fourier space (with
all modes above $n \Lambda$ set to zero); the same is done with the
Lagrangian displacement field $\v{s}$ which is the full displacement field constructed at the
relevant order in Lagrangian perturbation theory (this includes the curl component as well).
The displacement proceeds by using weighted mass elements, or ``particles.'' This method ensures that no noise is generated in the Eulerian-space fields on large scales, since mass is exactly conserved in the displacement process. Specifically, 
one runs over the regular Cartesian grid of $(N_g^{\rm CIC})^3$ ``particle positions'' $\vq_p$, assigning each particle a weight, or mass, given by the operator $O^{\rm L}$ at $\vq_p$, and depositing the mass at the Eulerian position of the particle,
\be
\vx_p = \vq_p + \v{s}(\vq_p,\tau).
\ee
For this deposition, we choose a cloud-in-cell scheme. More precisely, we choose
a CIC grid with $N_g^{\rm CIC} = N_g^{{\rm CIC},g}$, where $N_g^{{\rm CIC},g}$ is the grid size used in the construction of the halo density field, i.e. the data $\delta_g$, which are likewise assigned using a CIC scheme. For all results in this paper, $N_g^{\rm CIC} = 512$. \reffig{flowchart} summarizes the procedure schematically. We refer to \cite{paper_nLPT} for further implementation details.

Instead of displacing the linear-order operator $\tr [\v{M}^{(1)}]$, we simply add the Eulerian matter density, obtained by displacing a trivial weight field equal to unity, to the set of bias operators; in terms of the general bias expansion, both procedures are equivalent, but the latter allows for a simpler interpretation of the corresponding bias parameter. The final result is our set of Eulerian operators $\{ O(\vx,\tau)\}$.

Before continuing, we discuss the size of the grid on which the $\v{M}^{(n)}$ and $O^{\rm L}$ are constructed; the grid should have a sufficiently large Nyquist frequency to ensure that
all mode couplings are incorporated without aliasing (note that this is larger
than the grid size for the likelihood, which is determined by the condition
\refeq{NgL} above).
If we were only interested in modes $\v{M}^{(n)}(\vk)$ with $|\vk| \leq \L$,
then it would be sufficient to require $k_{\rm Ny} > (n+1)\L/2$ \cite{Michaux:2020yis,paper_nLPT},
since aliasing affects modes with wavenumbers greater than $|2 k_{\rm Ny}-n\L|$,
and we just need this to be greater than $\L$. 
However, since we still need to displace the operators into Eulerian space,
which couples modes with $|\vk| > \L$ to final modes below $\L$, we ensure
that none of the modes on the grid are affected by aliasing, which
requires $k_{\rm Ny} > n \L$. We choose the smallest even number of grid points
$N_g$ that satisfies this condition as our grid size.

\begin{table}[t]
\centering
\begin{tabular}{cccc}
\hline
\hline
Order & Leading bias operators & Higher-derivative operators & Total number of operators\\
\hline
$o=3$ & [7, \refeq{LagrBasis}] & $\lapl\d$ & 8\\
$o=4$ & [15, \refeq{LagrBasis}] & $\lapl\d$, $(\vn\d)^2$, $\lapl O^{(2)}$ [4] & 19 \\
$o=5$ & [29] & [13] & 42 \\
\hline
\hline
\end{tabular}
\caption{Number of relevant operators at each order, following \refeq{order} and \refeq{par}. The numbers in brackets give the total number of operators in each case. $O^{(2)}$ stands for the two second-order bias operators (second line in \refeq{LagrBasis}, but after displacement to Eulerian space).}
\label{tab:nO}
\end{table}

\subsection{Higher derivatives and ordering of operators}
\label{sec:hderiv}

The bias expansion is an expansion both in orders of perturbations, as considered above, and in derivatives. In our construction, we add higher-derivative operators iteratively to the set of Eulerian bias operators. These derivatives could equivalently be added in Lagrangian space. However, since the displacement is the most costly operation, it is more efficient to generate these new terms after the displacement. For each pair of operators $O,O'$ in the set of Eulerian operators, we add
\be
\lapl_x O, \quad \vn_x O' \cdot\vn_x O , \quad  O' \lapl_x O \ (\mbox{if } O \neq O')
\ee
to the set of operators. This set is chosen to be linearly independent (hence we exclude $O\lapl_x O$), and to capture a majority of higher-derivative operators. It does not capture the complete set of higher-derivative operators at second and higher order in perturbations however (see \cite{paper_nLPT} for details). We then repeat this application of derivatives recursively until all relevant operators are included.

The relevance of a given operator $O$ which starts at $n$-th order in perturbations, involves $2m$ derivatives, and $k$ stochastic fields is given by \cite{cabass/schmidt:2019}
\be
\epsilon(O) = \left(\frac{\L}{\knl}\right)^{n (3+\nlin)/2}  (\L R_*)^{2m} ( P_\eps^{\{0\}} \L^3)^{k/2},
\label{eq:order}
\ee
where $\nlin\equiv d\ln P_L(k)/d\ln k\Big|_{k=\L}$ is the linear power spectrum slope at the cutoff $\L$. 
The index $k$ is either 0 (for operators appearing in $\dgdet$) or 1 (for those appearing in the variance). 

Specifically, we determine the minimum relevance by selecting a value of $o$ as the maximum order of operators with no additional derivatives appearing in $\dgdet$, and then include all higher-derivative operators that have the same or higher relevance in $\dgdet$. Similarly, one would include all operators that, for $k=1$, have the same or higher relevance in the variance. As mentioned in \refsec{like}, we do not include stochastic operators for results in this paper however. We will show results for $o=3,4,5$.

In order to be able to easily compare results at different values of the
cutoff $\L$ and redshift $z$, we evaluate \refeq{order} at fixed parameter values, namely
\be
z=0;\quad \knl=0.25\iMpch;\quad \L = 0.14\iMpch;\quad
R_* = 5 \Mpch.
\label{eq:par}
\ee
The value of $R_*$ is a reasonable compromise given the Lagrangian radii of the
halo samples considered, while $\L=0.14\iMpch$ represents the middle of the range
in cutoff values we will consider below.
With these values, we obtain the sets of relevant operators given in \reftab{nO}. In case of $o=5$, we only list the total number of operators. Clearly, they multiply rapidly toward higher order.
Notice that the non-Gaussianity of the noise field, which we neglect in the
Gaussian likelihood of \refeq{lnPR}, only becomes formally relevant at $o=6$
\cite{cabass/schmidt:2019}. 

\section{Likelihood implementation details}
\label{sec:impl}

We now describe the numerical details of the implementation of the
real-space likelihood on a grid. Following the likelihood derived in \cite{cabass/schmidt:2020},
we generalize \refeq{lnPR} to include
a position-dependent variance $\sigma^2(\vx)$.
As input, we take the data $\d_g(\vx)$, which in our case is obtained from
assigning a rest-frame halo catalog at a given redshift, and the set of Eulerian
bias operators $O(\vx)$ constructed from the fixed initial conditions $\d_{\rm in}$
(but varying \se, which is implemented as described below). 
The steps for the computation of the likelihood thus start
from the boxes labeled with $\{O\}$ and $\d_g$ in the flowchart \reffig{flowchart}. 
The operator fields, combined with the bias parameters $b_O$, yield the deterministic field
$\dgdet[\{b_O\}; \se; \d_{\rm in}]$. The set of nuisance parameters consists of the $b_O$ as well as the
parameters entering the variance $\sigma^2$; for this paper, this is only a single
parameter $\sigma_0$, as discussed below.

As discussed in \refsec{like},
we first reduce all grids, i.e. $\d_g(\vx),\,\dgdet(\vx),\,\sigma(\vx)$,
from the base grid resolution $N_g$ to $N_g^{\rm red}$, where the Nyquist
frequency of the reduced grid matches the cutoff $\L$ (\refeq{NgL}). This reduction is
done in Fourier space and requires some care
to ensure the modes are properly mapped on the Nyquist planes
(see \refapp{reducesharpk}). While in Fourier space, we also set the $\vk=\v{0}$ mode
in each operator as well as the data to zero, ensuring that all fields have
vanishing mean. After the reduction, we then perform an
inverse Fourier transform on the reduced grids, and evaluate the likelihood
in real space:
\ba
-2 \ln\P(\d_g | \{b_O\}, \sigma_0; \se; \d_{\rm in}) =& \sum_{\vx}^{N_g^{\rm red}}
\frac{\left(\d_g(\vx) - \dgdet[\{b_O\}; \se; \d_{\rm in}](\vx)\right)^2}{[\sigma(\vx)]^2}  \vs
& + \frac{N_{\rm modes}}{(N_g^{\rm red})^3}
\sum_{\vx}^{N_g^{\rm red}} \ln \left( 2\pi \frac{N_g^6}{(N_g^{\rm red})^3} [\sigma(\vx)]^2 \right).
\label{eq:lnPRimp}
\ea
Notice that no explicit cutoff is necessary in the likelihood, since only modes with
$|k_i| \leq \L$ are represented on the grid. 
The normalization requires some explanation. First, we scale it from the
number of real-space grid points $(N_g^{\rm red})^3$ to the actual number
of independent modes $N_{\rm modes}$ computed as described in \refapp{reducesharpk}. Further, we add a rescaling factor $N_g^6/(N_g^{\rm red})^3$, with which
the log-likelihood for a constant $\sigma$ field returns the same value
it would when evaluating the likelihood in Fourier space on a grid of size
$N_g$. This is mostly done in order to cross-check the equivalence with the
Fourier-space formulation; one can equivalently set $N_g = N_g^{\rm red}$ in this term
without any impact on the inference, as it is an additive constant.

\refeq{lnPRimp} is straightforward to evaluate, however still explicitly
depends on the bias parameters, which requires one to search for a maximum
in a high-dimensional parameter space. As shown in \cite{paperII} and \cite{cabass/schmidt:2020}, it is possible to analytically marginalize over the bias parameters;
this is because the log-likelihood \refeq{lnPRimp} is a quadratic polynomial in the
bias parameters (they enter linearly in $\dgdet$). 
In the case that all bias parameters are marginalized over (in the notation of the above references, $\mu\to 0$), the likelihood becomes
\ba
-2 \ln\P(\d_g |\sigma_0; \se;\d_{\rm in}) &= C - \sum_{O,O'} B_O(A^{-1})_{O O'}B_{O'} + \ln \det A\vs
&\quad
+ \frac{N_{\rm modes}}{(N_g^{\rm red})^3}
\sum_{\vx}^{N_g^{\rm red}} \ln \left( 2\pi \frac{N_g^6}{(N_g^{\rm red})^3} [\sigma(\vx)]^2 \right)
+ \ln \det {\rm C}_{\rm prior} 
\,,
\label{eq:lnPRmarg}
\ea
where
\begin{subequations}
\label{eq:margdefs}
\begin{align}
C(\sigma_0) &= \sum_{\vx}^{N_g^{\rm red}}\,\frac{1}{[\sigma(\vx)]^2} \big(\d_g(\vx)\big)^2 \label{eq:margdefs-1} \\
B_O(\sigma_0; \se; \d_{\rm in}) &= \sum_{\vx}^{N_g^{\rm red}}\,\frac{\d_g(\vx) O(\vx)}{[\sigma(\vx)]^2} + \sum_{O'} 
({\rm C}_\text{prior}^{-1})_{OO'} b^{\rm prior}_{O'} \label{eq:margdefs-2} \\
A_{OO'}(\sigma_0; \se; \d_{\rm in}) &= \sum_{\vx}^{N_g^{\rm red}}\,\frac{O(\vx) O'(\vx)}{[\sigma(\vx)]^2} 
+ ({\rm C}_\text{prior}^{-1})_{OO'}\,\,, \label{eq:margdefs-3}
\end{align}
\end{subequations}
while $b^{\rm prior}_O$ and $C_{\rm prior}$ denote the mean and covariance of a Gaussian prior on the bias parameters. While the code implementation allows for priors, for this paper we drop the prior terms, i.e. formally send $C_{\rm prior}^{-1} \to 0$, corresponding to a uniform prior on the bias parameters. Note that the $O(\vx)$, and hence $B_O$ and $A_{OO'}$, depend on \se and $\d_{\rm in}$ via the forward model. In this paper, we always show results marginalized over all $b_O$ (while Refs.~\cite{paperII,paperIIb} did not marginalize over $b_1$). 

All the grid operations are straightforwardly parallelized (using OpenMP in our implementation).
For the matrix operations (inverse and determinant), we use the LU decomposition with full column pivoting as provided by the Eigen C++ library \cite{eigen}.\footnote{The matrix $A_{OO'}$ is positive definite and as such lends itself to a Cholesky decomposition. However, we have found this to be less accurate than the LU decomposition.} Specifically, we write
\be
\sum_{O,O'} B_O(A^{-1})_{O O'}B_{O'} = \v{B}^\top \cdot \v{X} \quad\mbox{where $\v{X}$ satisfies}\quad \v{A}\cdot\v{X} = \v{B},
\ee
avoiding the explicit computation of the matrix inverse.

The computation of the \se profile likelihood proceeds by finding the maximum of the likelihood \refeq{lnPRmarg} over all free parameters for a range of \se values \cite{paperII}. Specifically, we determine the maximum log-likelihood for the values
\be
\Ain \equiv \frac{\se}{\se^{\rm fid}} \in \{ 0.9,\ 0.95,\ 0.98,\ 1.00,\ 1.02,\ 1.05,\ 1.1 \} ,
\ee
which yields the profile likelihood $P_{\rm prof}(\Ain)$.
The different values of $\Ain$ are implemented by rescaling the fiducial linear density field used to generate the initial conditions of the N-body simulations by the factor $\Ain$ \emph{before} constructing the LPT forward model and bias operators; i.e., we use $\d_{\rm in}(\vx) = \Ain \d_{\rm in}^{\rm fid}(\vx)$ (hence the subscript on $\Ain$). In order to obtain a precise representation of $\d_{\rm in}^{\rm fid}(\vx)$, we have modified the initial conditions generator of Ref.~\cite{ic2lpt} (which is based on that of \cite{2005MNRAS.364.1105S}) to write the linear density field to disk, before it is used to generate the 2LPT particle displacements.
Again, unlike the results in previous papers in this series, the final matter density field comes out of the nLPT forward model and is not taken from an external code or
simulations. 
As described in \cite{paperII,paperIIb}, the maximum-likelihood value $\hAin$ and its error are determined through the maximum and curvature around the maximum of $P_{\rm prof}(\Ain)$.

In our case, where the initial phases and hence the bias operators constructed from them are fixed, the only free parameters remaining in \refeq{lnPRmarg} at fixed value of \se (equivalently $\Ain$) are those entering the variance: the constant variance parameter $\sigma_0$, corresponding to the square-root of the spatial average of $P_\eps(\vx)$, and one free coefficient for each relevant stochastic operator. 
This maximization is done using \textsc{Minuit} as implemented in the \textsc{root} package. 
Unfortunately, the profile likelihood is easily spoiled if significantly different maximum-likelihood values of $\sigma_0$ are found at different values of $\Ain$, for example due to numerical instabilities in the maximization; variations in $\sigma_0$ at the few-percent level are already sufficient to lead to significant noise in $\hAin$. While this issue is manageable for the $o=3$ expansion, it becomes progressively worse at higher orders. For this reason, we do not consider a field-dependent covariance here, but instead restrict the covariance to a constant, $\sigma^2(\vx) = \sigma_0^2$.
Notice that this issue should be alleviated once a full joint sampling of $\se$ and the stochastic parameters is performed, since then the likelihood is evaluated consistently at each point in the joint parameter space.
Apart from this numerical issue, we have found that the field-dependent covariance only has a minor impact on the \se inference. We discuss this in \refsec{sigmax}. 

Finally, in order to determine the maximum-likelihood value $\hAin$ and its error from the profile likelihood, we fit a quadratic polynomial to the log-likelihood, which yields $\hAin$ as the point of maximum and the estimated error as the inverse square root of the curvature.

\section{Results}
\label{sec:results}

We now present results for the maximum-likelihood value of \se (or equivalently $\AS$) inferred from rest-frame halo catalogs for fixed initial phases, phrased in terms of the maximum-profile-likelihood value $\hAin$. Unbiased inference corresponds to values of $\hAin$ that are consistent with 1 within errors. The default
halo catalogs are the same as those used in \cite{paperII,paperIIb},
and are described in \refapp{halos}; they consist of four sharp, disjoint
mass bins covering the range $10^{12.5}-10^{14.5} \Msunh$ at redshifts $0, 0.5, 1$, identified in two simulation realizations of $(2000\Mpch)^3$ volume each. The only difference to the samples reported on in previous work is that we now
use halos identified in N-body simulations with a starting redshift $z_{\rm in}=24$ instead of 99, as employed in the previous papers in this series. The
reason is discussed in \refsec{disc}. 

\begin{figure*}[thbp]
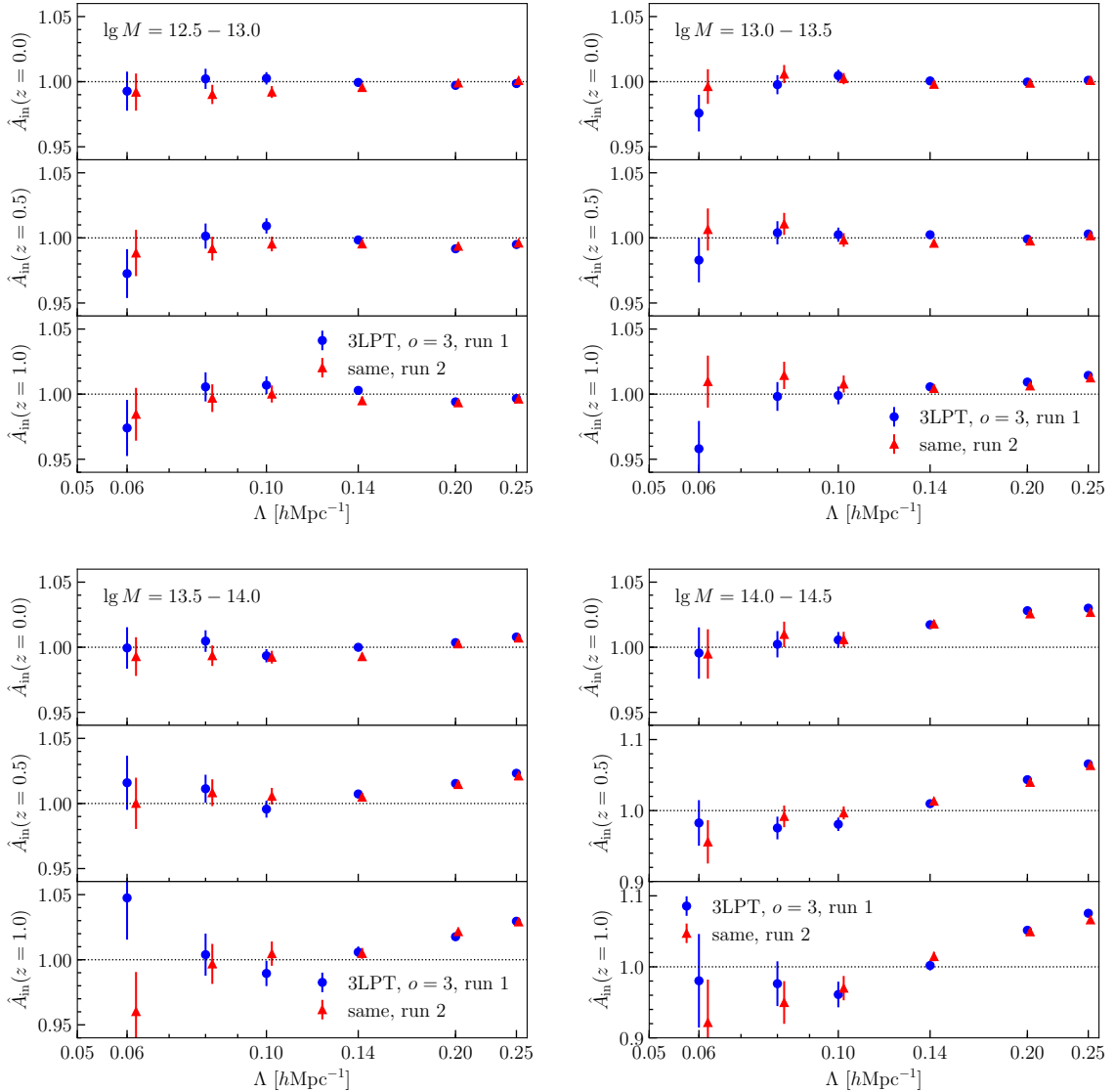
%
  \fourmassbinplots{sigma8_mlemargRealConst_3LPT3D_prod_run12}
  \cprotect\caption{
    Results on the estimated value of \se, relative to the ground truth, as a function of cutoff value $\L$. The different panels show the four mass bins considered (with $M$ in units of $\Msunh$), and each panel shows the results for three redshifts: $z=0,\,0.5,\,1$. The results shown are for 3LPT with an $o=3$ bias expansion, and for two simulation realizations. The scatter between the two simulation realizations is consistent with the statistical errors inferred from the profile likelihood.
    \label{fig:LPTprod_3d}}
\end{figure*}%

\reffig{LPTprod_3d} shows the results for the 3LPT forward model with a third-order bias expansion; more precisely, $o=3$ in the ordering described in \refsec{hderiv}. The reported error bars, which take into account both halo stochasticity and cosmic variance (see below), clearly grow for the more rare high-mass halo samples at higher redshifts.

The expected convergence to $\hAin = 1$ as $\L\to 0$ is seen for all masses and redshifts. Note
that the inferred \se value never differs from the truth by more than $\sim 4\%$ for all our samples and all cutoff values considered. Excluding the lowest cutoff value, which has the largest statistical error bars, as well as the highest cutoff value $\L=0.25\iMpch$, which is close to the nonlinear scale at $z=0$,  the \se inference is in fact accurate to 2\% or better for all samples for this third-order forward model.
The rate at which $\hAin$ diverges from 1 when going to larger $\L$ clearly depends on halo mass and redshift. We return to this point below when comparing results at different orders $o$.

\reffig{LPTprod_3d} shows results for both simulation realizations. Note that, since the phases are fixed to the true values used in the initial conditions for each simulation realization, the only source of cosmic variance between the two realizations is due to the different realizations of the modes \emph{above} the cutoff $\L$. Thus, at fixed cutoff value and halo sample, the results from different simulation realizations are sampled from the same underlying distribution.
The width of this distribution should, if the EFT likelihood is accurate, be correctly captured by the statistical error inferred from the profile likelihood, which is controlled by the effective noise amplitude $\sigma_0$. Note that the previous papers in this series \cite{paperI,paperII,paperIIb} incorrectly suggested that there would be an additional cosmic variance contribution to the error on $\hAin$ beyond that given by the profile likelihood.

Indeed, the scatter between the results from the two realizations in \reffig{LPTprod_3d} is consistent with the statistical error bar inferred from the profile likelihood. We have verified this by histogramming the quantity $(\hAin({\rm run\ 1}) - \hAin({\rm run\ 2}))/\sqrt{\sigma^2({\rm run\ 1})+\sigma^2({\rm run\ 2})}$ for the results shown in \reffig{LPTprod_3d}, i.e. combining all cutoff values and halo samples. The resulting distribution has an RMS of $1.2\pm 0.1$, indicating that, in addition to unbiased inference of $\hAin$ on large scales, the EFT likelihood also correctly estimates the error on $\hAin$.

\textbf{Comparison to Fourier-space likelihood:} before continuing,
it is worth comparing the results of the real-space likelihood to that
of the Fourier-space likelihood employed in \cite{paperII,paperIIb},
using the same Lagrangian forward model and bias expansion for both likelihoods 
in order to restrict the comparison to the likelihood itself.
This is shown in \reffig{LPTprod_2dF} in \refapp{figs}. 
Note 
that the real-space likelihood employs a cubic $k$ cut, while the Fourier-space
implementation of \cite{paperII,paperIIb} employs a spherical cut. Hence,
the modes used in each case are not the same, and we do not expect exact agreement.\footnote{We have verified that, when restricting to precisely the same modes and a constant covariance in both cases, the results from both likelihoods agree precisely as expected following the discussion in \refsec{like}.}
In fact, the real-space likelihood employs modes with slightly larger wavenumber, up to $\sqrt{3} \L$ as compared to $\L$ for the spherical cut; the effect of this is visible for the lowest values of $\L$, where the error bars in the Fourier-space likelihood results are noticeably larger than the corresponding real-space ones. Given these differences, we find very good agreement.

\begin{figure*}[thbp]
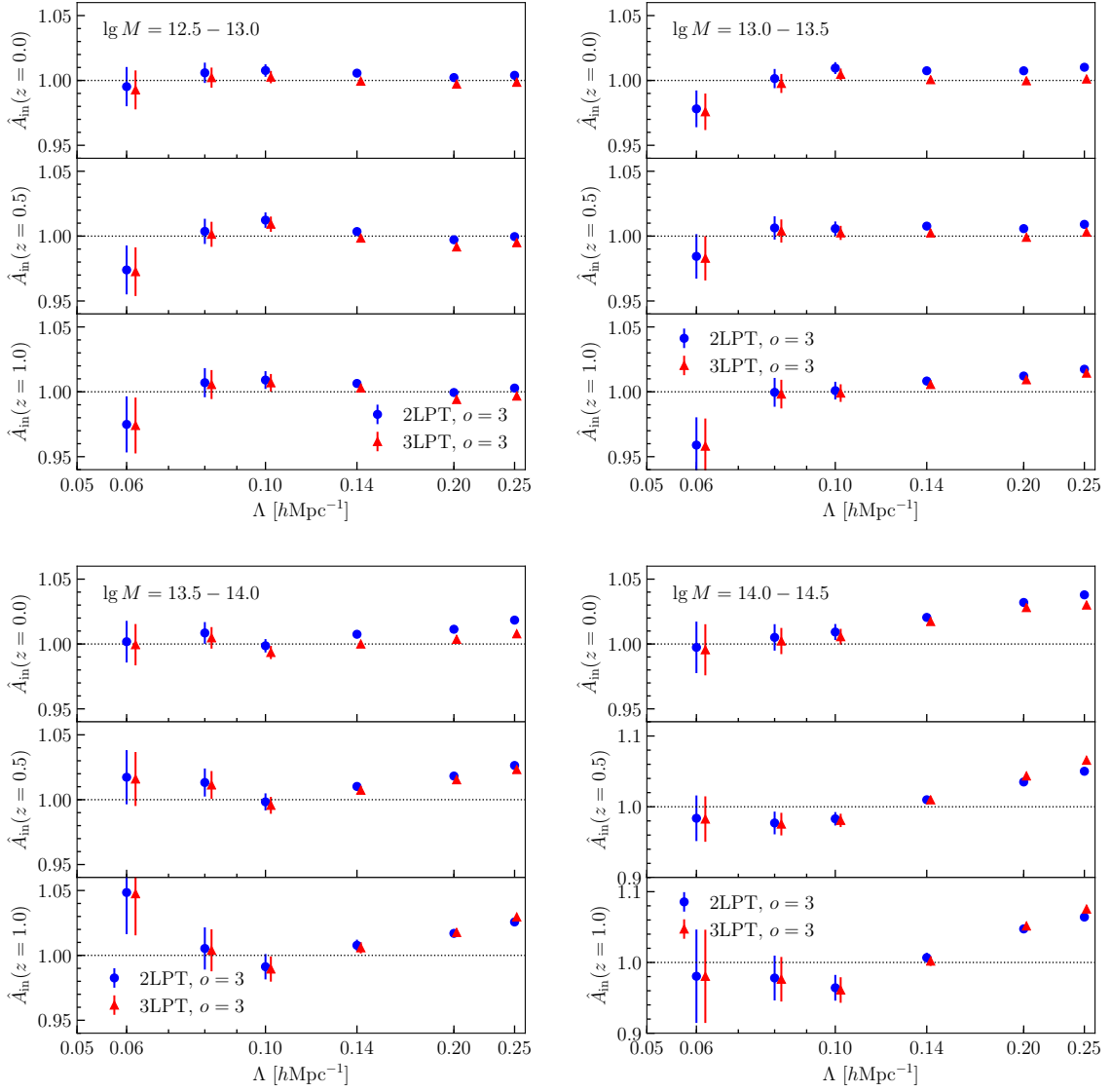
%
  \fourmassbinplots{sigma8_mlemargRealConst_2LPT3D_prod}
  \cprotect\caption{Same as \reffig{LPTprod_3d}, comparing results using 2LPT and 3LPT forward models, both with an $o=3$ bias expansion. Results for run 1 are shown here and in all following figures.
    \label{fig:LPTprod_2d}}
\end{figure*}%

\begin{figure*}[thbp]
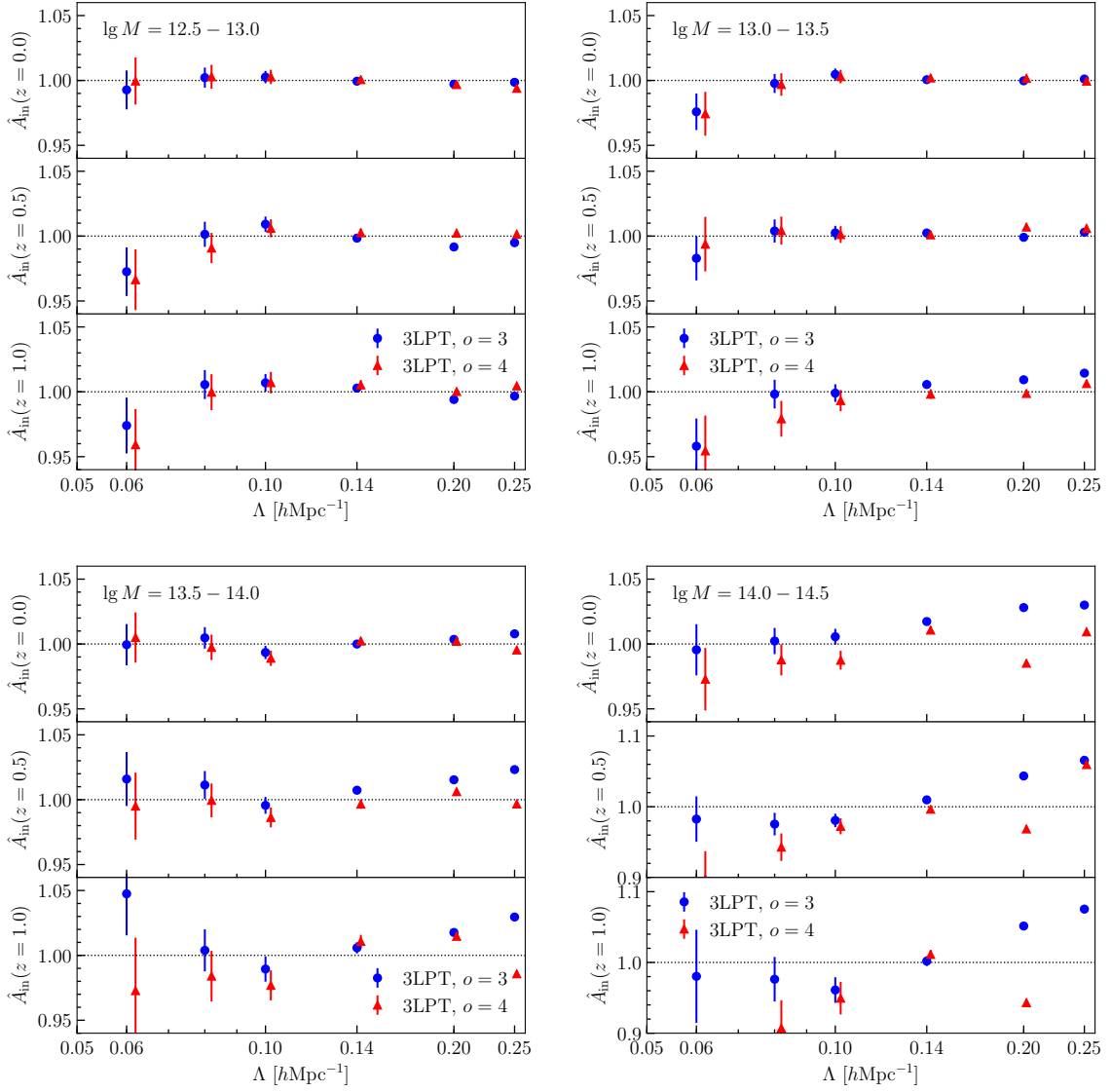
%
  \fourmassbinplots{sigma8_mlemargRealConst_3LPT3D_prod}
  \cprotect\caption{
    Same as \reffig{LPTprod_3d}, comparing results using 3LPT forward model, with an $o=3$ vs $o=4$ bias expansion.
    \label{fig:LPTprod_3d_vs_4d}}
\end{figure*}%

\textbf{Effect of LPT order:} \reffig{LPTprod_2d} compares results
of second- and third-order LPT (2LPT and 3LPT, respectively), both for the $o=3$ bias expansion.
The results are similar, but since 3LPT yields more power in the
density field and displacement, the estimated value of
$\hAin$ generally moves down slightly when compared to 2LPT. In most cases, this moves $\hAin$ closer to unity.

\textbf{Effect of bias order:} \reffig{LPTprod_3d_vs_4d} compares results
for 3LPT with the $o=3$ (as in \reffig{LPTprod_3d}) and the $o=4$ bias expansions. For cutoffs $\L \leq 0.1\iMpch$, the fourth-order bias terms do not change the $\se$ inference, except for the rarest halo samples where they actually appear to lead to increased scatter; note that the $o=4$ bias expansion marginalizes over 19 parameters, compared to 8 for $o=3$ (\reftab{nO}). For higher cutoff values, the $o=4$ case does perform somewhat better, essentially increasing the reach of the forward model of the halo density field to smaller scales.
Similar conclusions hold when going to even higher order, $o=5$, which increases the number of free parameters by another factor of 2.
This trend becomes even clearer when plotting $\hAin$ at fixed $\L$ for the different halo samples.

\clearpage

\begin{figure*}[thbp]%
  \centerline{\resizebox{\hsize}{!}{
      \includegraphics*{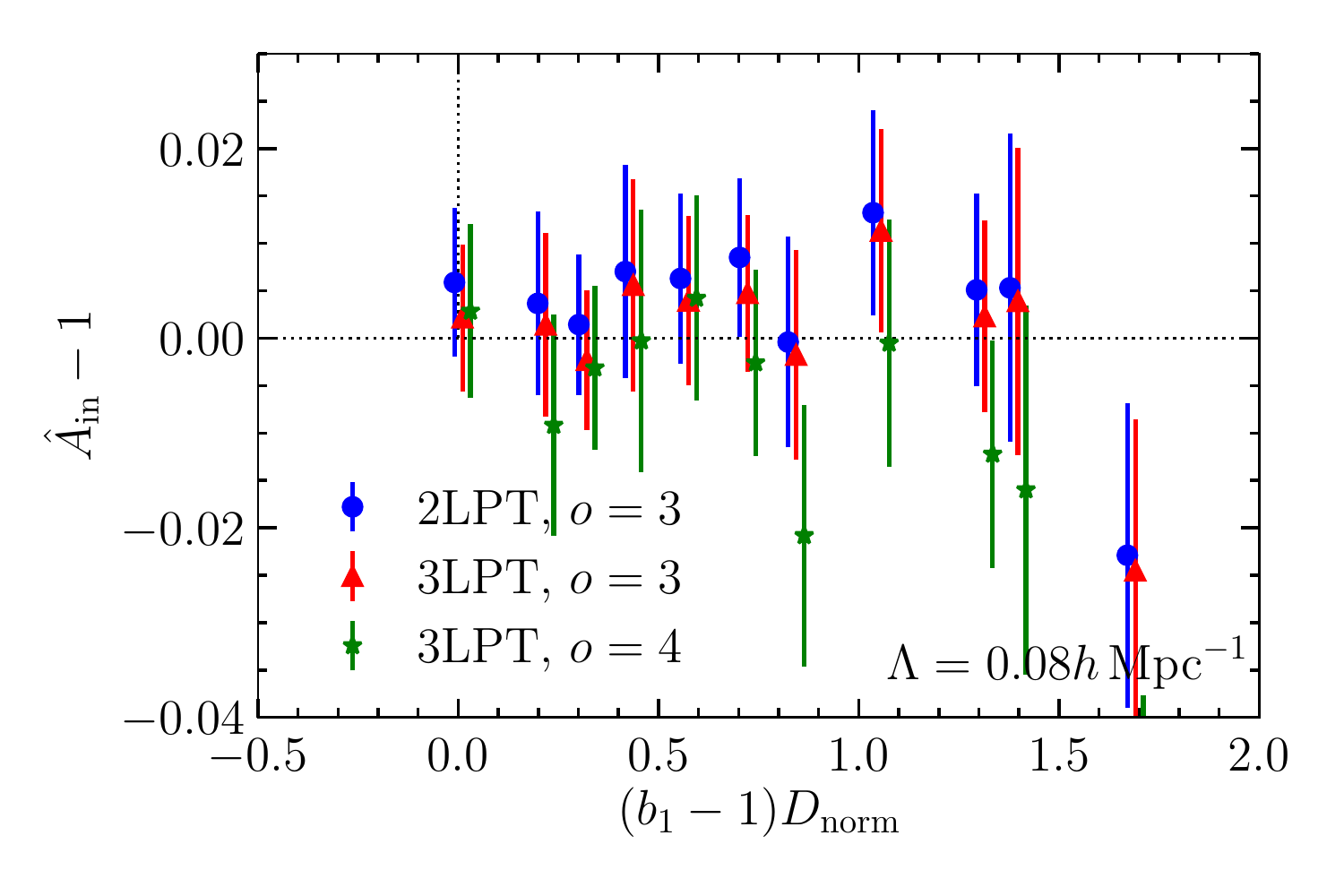}
      \includegraphics*{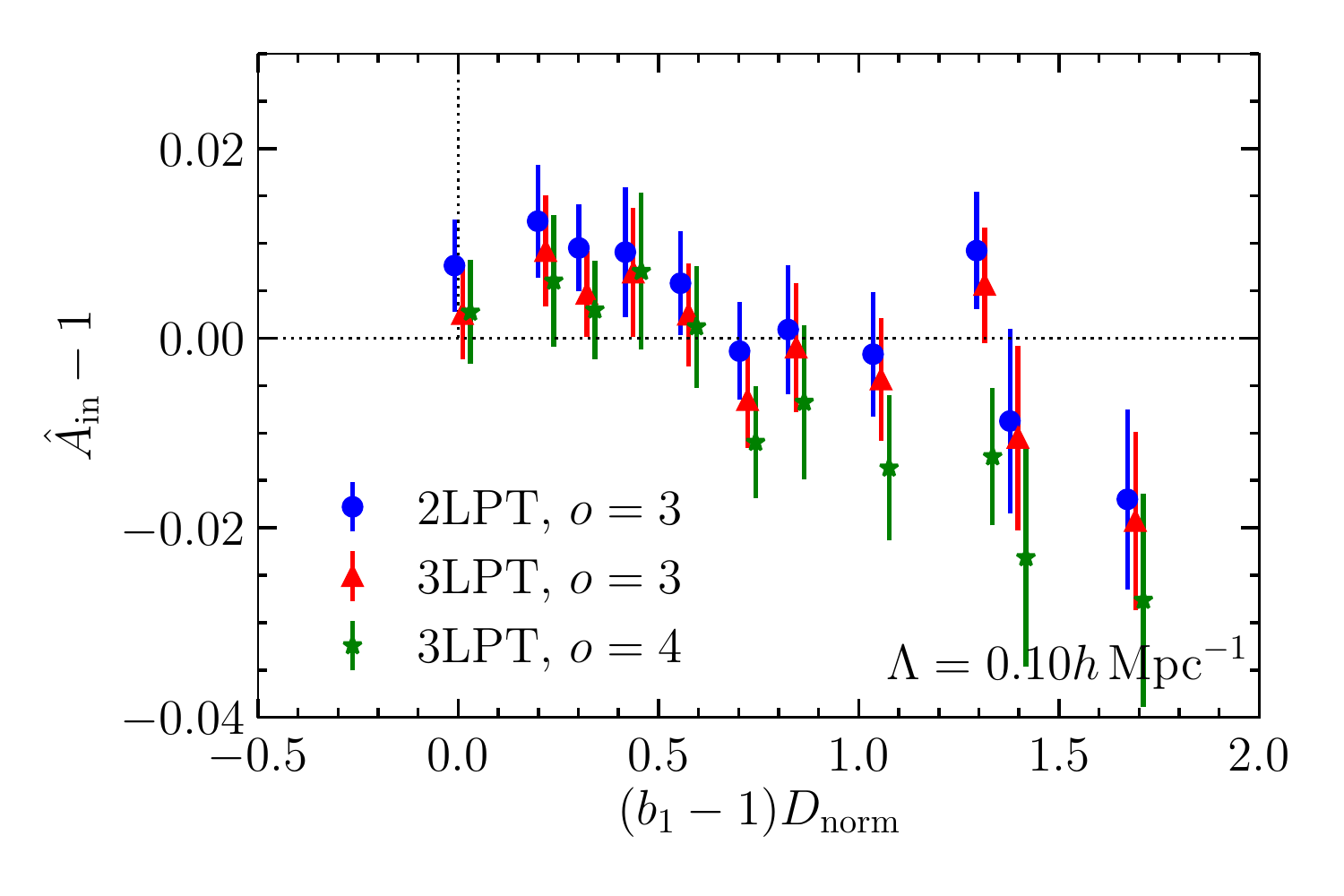}
      }}
  \centerline{\resizebox{\hsize}{!}{
      \includegraphics*{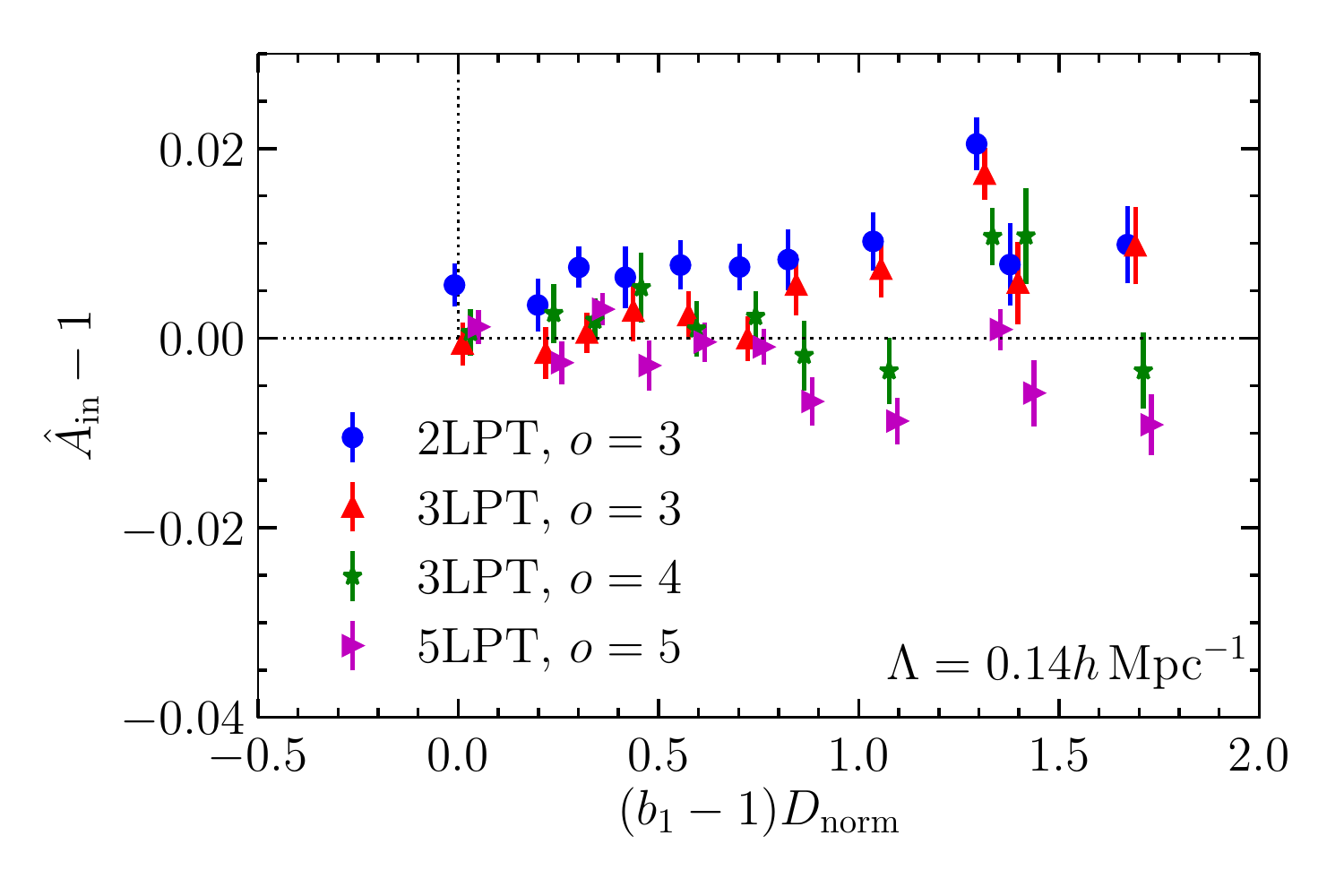}
      \includegraphics*{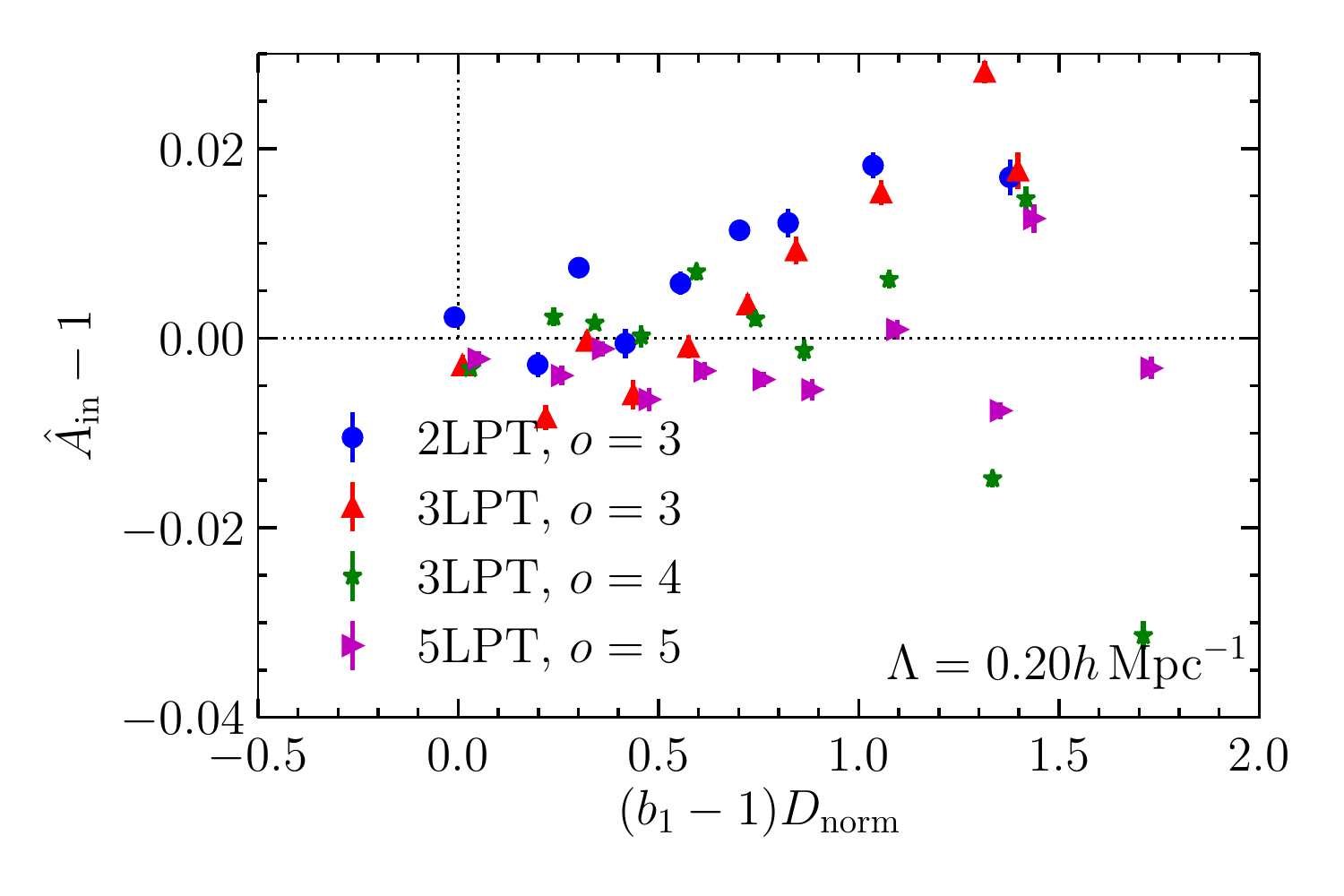}
      }}
  \cprotect\caption{Maximum-likelihood value $\hAin-1$ for all halo samples and redshifts (but for one simulation realization only) at a fixed cutoff value. The x axis shows the combination $(b_1-1) D_{\rm norm}$, where $D_{\rm norm} = D(z)/D(0)$ is the normalized growth factor at the redshift of the given sample. The different panels show different cutoff values as indicated. In each panel, we show results for different forward model/bias expansions. This gives an overview of the overall performance of different expansions at different cutoff values. 
    \label{fig:alpha_vs_X}}
\end{figure*}%

\textbf{$\bm{\Ain}$ vs. bias:} So far, we have discussed $\hAin$ as a function
of cutoff $\L$ for individual halo samples. \reffig{alpha_vs_X} shows an
alternative representation, where \emph{all halo mass bins and redshifts are plotted
in a single panel, but at fixed cutoff.} This gives a good overview of the performance
of a given expansion order at a fixed cutoff. We choose to plot results as
the fractional deviation of the inferred \se value from the truth, i.e. $\hAin-1$,
as a function of the combination $b_1 D_{\rm norm}$ where $b_1$ is the linear
bias and $D_{\rm norm} = D(z)/D(0)$ is the normalized growth factor at the redshift of the given halo sample. As argued in \cite{paperIIb}, $b_1 D_{\rm norm}$
is a rough indicator for the magnitude of higher-order bias contributions
(that is, higher order in perturbations rather than derivatives).
Since we marginalize over $b_1$ here, we adopt the values for $b_1$ reported in
\cite{paperIIb} for the same halo samples using the third-order likelihood; this is entirely sufficient
for this purpose. The different panels in the figure show different cutoff
values. Some interesting trends can be gleaned from this representation:
\begin{itemize}
\item For $\L=0.08\iMpch$, all results are consistent with $\hAin=1$ within errors; no significant improvement is seen for higher LPT or bias orders.
\item At $\L=0.1\iMpch$, deviations start to become statistically mildly significant for the most highly biased samples, in agreement with the conclusions of \cite{paperIIb}.
\item For $\L=0.14\iMpch$, the deviations are now significant statistically, albeit not much larger in magnitude; for this cutoff value, going from 2LPT to 3LPT, and from $o=3$ to $o=4$ bias expansions each reduce the bias in \se significantly. The results for $o=5$ do not generally improve upon those with $o=4$ for the more highly biased samples, likely because the fifth-order contributions are still small on those scales.
\item $\L=0.2\iMpch$: Similar conclusions hold as for $\L=0.14\iMpch$, except that the $o=5$ bias expansion now marginally improves the residuals over $o=4$ as well.
\end{itemize}
  
To summarize, we find the expected reduction in the systematic bias on the inferred \se value when lowering $\L$ at fixed bias order, or when increasing the bias order at fixed $\L$. The exception is that for $o > 3$, some instabilities appear at lower values of $\L$, especially for the rare halo samples. It would be interesting to revisit this issue with full sampling instead of the profile likelihood. In any case, we find that in all cases where these instabilities appear,  a lower-order bias expansion is sufficient to yield unbiased results to within errors (e.g., $o=3$ for $\L =0.1\iMpch$, or $o=4$ for $\L = 0.14 \iMpch$). Overall, it appears that not much improvement is obtained when going beyond $o=4$. 

It is also worth noting that the inferred statistical errors on $\hAin$ at
fixed $\L$ do not grow significantly when going to higher orders in the expansion, despite the additional free parameters that are being marginalized over.

\section{Discussion}
\label{sec:disc}

We now discuss investigations on issues apart from the expansion order
and halo mass and redshift presented above.

\subsection{Position-dependent variance}
\label{sec:sigmax}

\begin{figure*}[thbp]
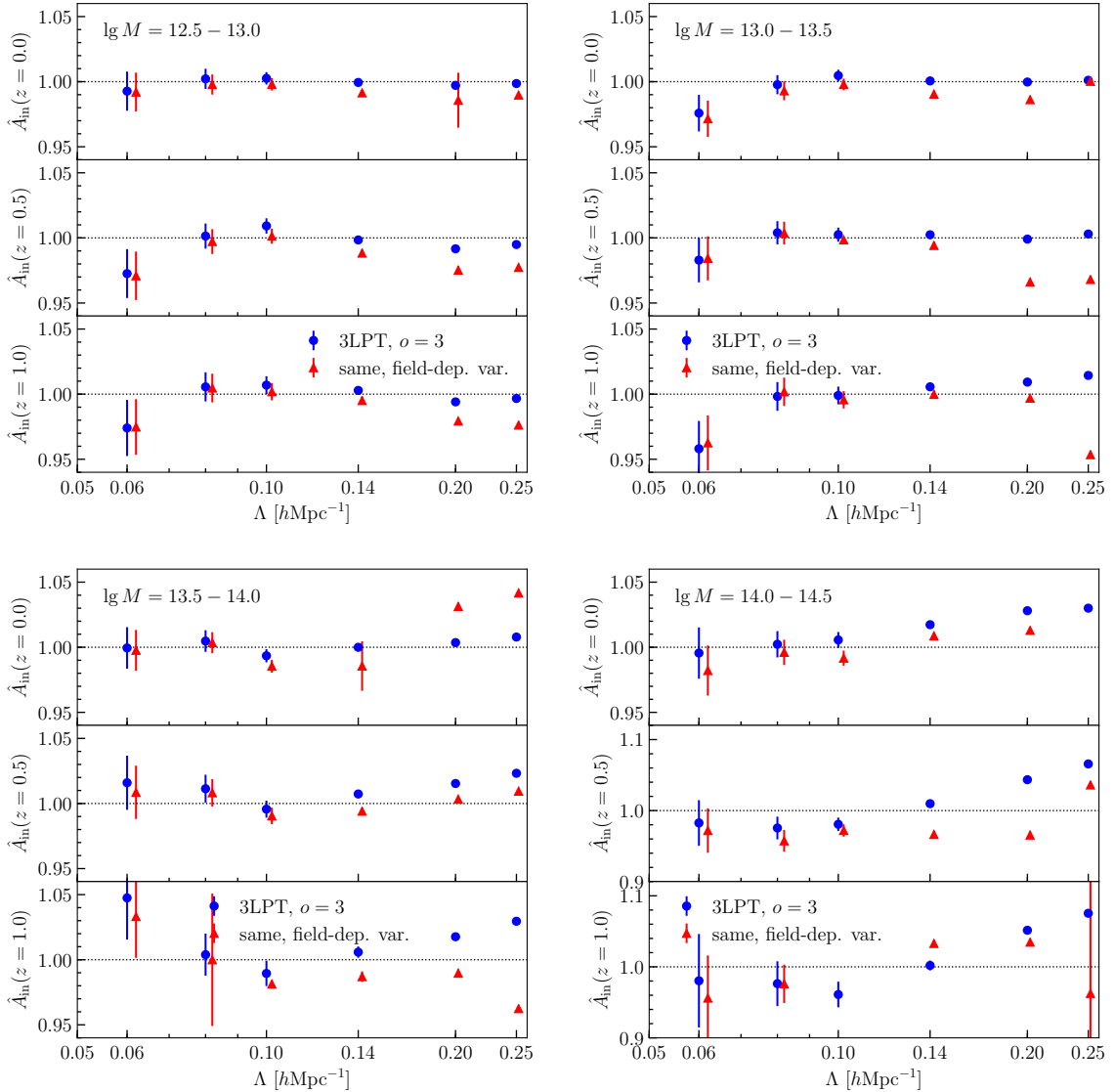
%
  \fourmassbinplots{sigma8_mlemargRealConst_vs_General_3LPT3D_prod}
  \cprotect\caption{Effect of allowing for a position-dependent variance, in case of the $o=3$ bias expansion.
    \label{fig:LPTprod_margrealgeneral}}
\end{figure*}%

The results shown so far are based on a constant variance field $\sigma^2(\vx) = \sigma_0^2$.
At the order in perturbations that we work in however, field-dependent terms
in the variance formally become significant.
The effect of the density-dependent variance on the inference is essentially
to upweight regions of less noise, and downweight regions with higher noise.
Thus, even if the density-dependent variance is formally relevant in perturbation
theory, we expect the constant-variance case to be merely suboptimal, but not 
necessarily biased.

\reffig{LPTprod_margrealgeneral}
compares results allowing for a position-dependent variance including the
relevant terms at order $o=3$. Specifically, the variance field is
written as the square of
\be
\sigma(\vx) = \sigma_0 \left[1 + \sum_O r^\eps_O O(\vx)\right] ,
\label{eq:sigmax}
\ee
where the sum runs over all operators that are relevant for the case
considered (using \refeq{order} with $k=1$).

We find that for moderately biased samples, the results are compatible in most cases,
although the constant-variance results are generally less biased. 
More significant differences are visible for the more highly biased samples.
However, the results with position-dependent variance are overall less
stable than the constant-variance case. 
We attribute this to the numerical issues discussed in \refsec{impl}; indeed,
the profile likelihoods show several outliers for these highly biased cases,
for which the value of $\sigma_0$ differs significantly between neighboring
values of $\Ain$, which is not expected physically. 
It is clearly worth revisiting this issue using a full sampling approach.

\subsection{Beyond the Einstein-de Sitter approximation}
\label{sec:BEdS}

We now turn to the impact of the EdS approximation. As discussed in \refsec{lptbias}, we do not consider additional bias terms induced beyond EdS, which appear at fourth order in perturbations. However,
the LPT implementation incorporates general expansion histories \cite{paper_nLPT}, so we can
determine the effect of going beyond the EdS approximation in LPT itself,
which formally appears already at second order in perturbations, where
$D_2 \neq D^2$. Note that we always use the linear growth factor $D$ for the
$\L$CDM simulation cosmology. 
This comparison is shown in \reffig{LPTprod_3d_BEdS} in \refapp{figs}. Both cases agree to
within fractions of a percent. We conlude that the EdS approximation is
not a significant source of systematic uncertainty at this level.

\subsection{N-body accuracy}
\label{sec:Nbody}

In any study involving simulation results, a quantification of the
numerical error in the simulations is in order. Here, we are only concerned with numerical issues that could
affect the \se inference; this also includes the halo finder used to
construct the halo sample (we turn to this in the next section). Given the general bias expansion that the forward model is built on,
any issue that can be captured by small-scale noise and error in the force
calculation should not lead to a bias in \se, as all local effects are captured
by bias terms and the noise covariance.

Following this reasoning, the N-body aspects of most concern are the accuracy
of the time integration, as this could potentially affect the accurate
calculation of the growth of the large-scale perturbations, and transients
from the initial conditions. We have investigated the former effect
by performing a re-simulation of our fiducial realization with increased
accuracy: specifically, we changed the Gadget2 parameters as follows:
\ba
\mbox{\texttt{ErrTolIntAccuracy}:}&\quad  0.025 \to 0.01 \vs
\mbox{\texttt{MaxRMSDisplacementFac}:}&\quad  0.2 \to 0.1 \vs
\mbox{\texttt{MaxSizeTimestep}:}&\quad  0.025 \to 0.01 \,.\nonumber
\ea
We have not found even a slight effect on the \se inference, concluding
that the standard Gadget2 precision settings employed in our fiducial
simulations are entirely sufficient at the percent level.

\begin{figure*}[thbp]
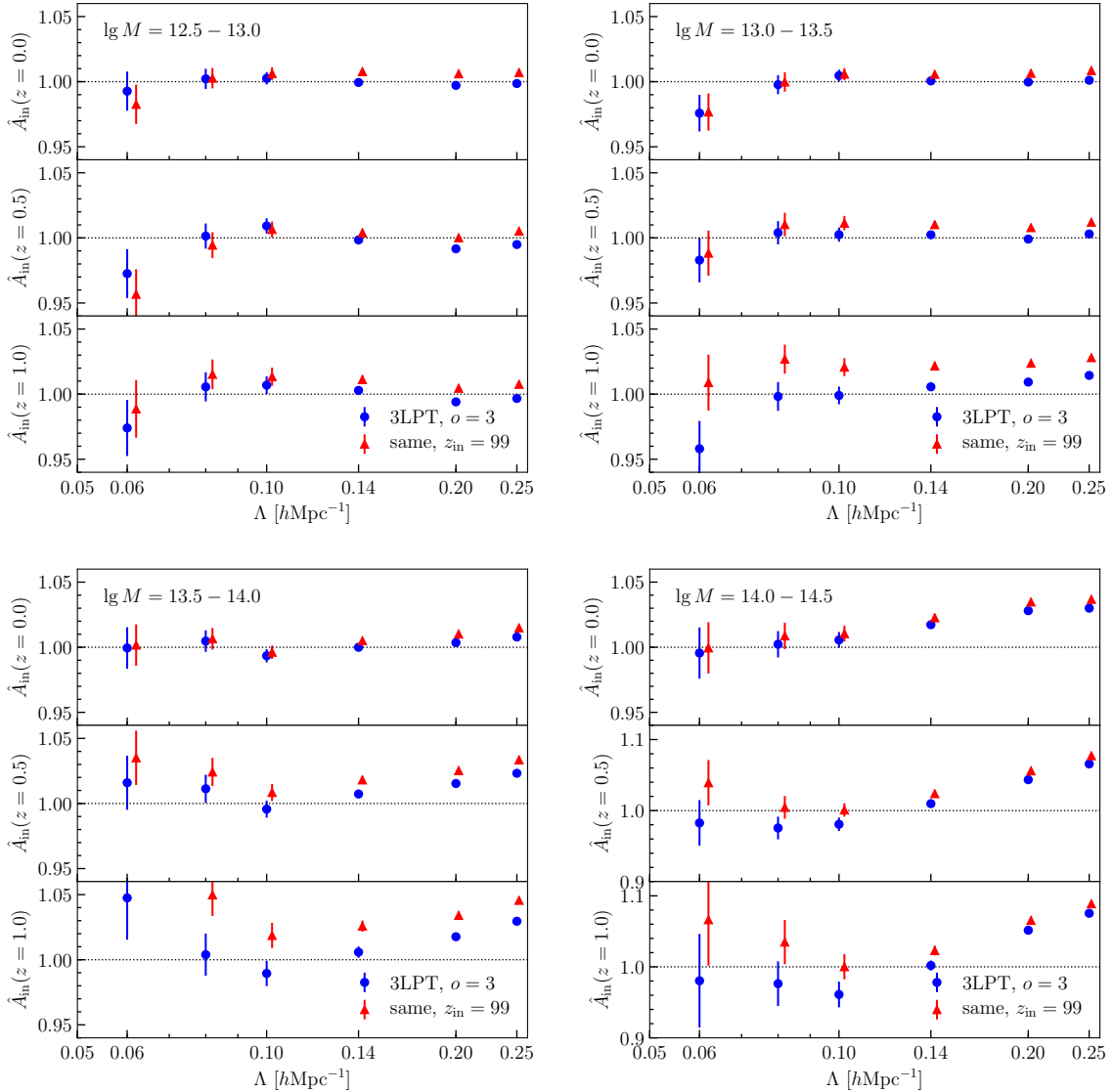
%
  \fourmassbinplots{sigma8_mlemargRealConst_3LPT3D_zin}
  \cprotect\caption{Effect of varying the initial redshift of the N-body simulations in which the halo samples are identified; 2LPT is used to initialize the simulations in both cases. The blue points show the fiducial case of $z_{\rm in}=24$ adopted in this paper, while red points show $z_{\rm in}=99$, corresponding to the simulations used in previous references.
    \label{fig:LPTprod_3d_zin}}
\end{figure*}%

The conclusions are quite different for transients from the initial conditions,
which in our case are always obtained from 2LPT, leaving the choice of starting redshift $z_{\rm in}$ for the N-body simulations. 
This is a subtle issue, as starting later (lower $z_{\rm in}$) leads to inaccuracies due to incorrect
nonlinear evolution, while starting earlier (higher $z_{\rm in}$) leads to
artefacts because of the grid on which the particles are initially placed
(``pre-initial conditions'') and the associated noise. While
the simulations initially performed for \cite{2017MNRAS.468.3277B}
were started at $z_{\rm in}=99$, we here choose a significantly later
starting redshift of $z_{\rm in}=24$ following the reasoning of
\cite{Michaux:2020yis,Nishimichi:2018etk}. Ideally, one would use a higher-order
LPT implementation to generate the initial conditions, but we refer this
to future work.\footnote{As the LPT forward model presented here is not distributed-memory capable, we cannot use it for generating the required 1536$^3$ particle grid.}

This choice is verified by the \se inference, as shown in \reffig{LPTprod_3d_zin}: there are percent-level shifts between $z_{\rm in}=24$ and $z_{\rm in}=99$
in some cases, with the former results being closer to unbiased.
A systematic investigation of the best choice of starting redshift is
deferred to future work. Clearly however, systematic errors due to
transients cannot be neglected when evaluating the accuracy of the \se
inference from perturbative approaches.

\subsection{Halo sample}
\label{sec:rockstar}

Following the discussion in the previous section, we do not expect the
\se inference to depend in any significant way on the halo finder and
mass definition employed, as long as the mass is determined from locally observable
quantities, such as the density (or distance to the nearest neighbor particle)
and relative velocity of a particle with respect to the halo center. Thus, the requirements
for a \se inference accurate to 1\% are much less stringent than, say, for
a 1\% measurement of the halo mass function.

As an illustration, 
\reffig{rockstar} (\refapp{figs}) shows the comparison of the \se results obtained from
a halo sample obtained from the \textsc{Rockstar} halo finder (see \refapp{halos} for details) and our fiducial set. No significant differences are found. 
We have also performed an analysis excluding subhalos from our fiducial sample.
We again find no significant difference to the fiducial case (only a modest fraction of our
fairly high-mass halos are subhalos).

\section{Conclusions}
\label{sec:conc}

We have presented a real-space implementation of the EFT likelihood,
which allows for the straightforward incorporation of observational effects
such as the survey window function. 
The EFT likelihood marginalizes precisely over those parts of the likelihood $\P(\d_g | \{ b_O, \sigma_0 \}, \se)$ of a biased tracer density field $\d_g$ that are
affected by nonlinear, spatially local but temporally nonlocal structure
formation. The real-space formulation was previously
presented in \cite{cabass/schmidt:2020}, but its actual implementation
involves some subtleties in the grid reduction (\refapp{reducesharpk}).
We presented
results on the inference of \se from a rest-frame halo catalog using the
real-space likelihood and a new Lagrangian forward model described in
detail in an upcoming paper \cite{paper_nLPT}. As in previous papers
in this series, we marginalize analytically over the substantial number of free
bias parameters (up to 42).

Our numerical results, presented in \refsec{results} and perhaps best
summarized by \reffig{alpha_vs_X}, fully show the expected convergence behavior as a function of scale, for the different expansion orders considered.
Compared to the recent Ref.~\cite{paperIIb}, which employed a Eulerian
bias expansion with an additional cutoff, the accuracy of the \se inference
has further improved. \bfem{For cutoff values $\L \leq 0.14\iMpch$, the residual
bias in \se is less than 2\%, and within 1\% for the majority of halo samples.}

Bias models beyond cubic order, $o=4$ and $o=5$,
further improve the \se inference over the cubic expansion for higher cutoff
values $\L > 0.14\iMpch$, in particular for the more highly biased samples.
Somewhat surprisingly, the bias in \se remains under control even for
a cutoff value of $\L=0.25\iMpch$, which approaches the nonlinear scale.
This is likely partially explained by the fact that the small scales
are already shot noise dominated for most of the halo samples;
we still generally find a reduction in the statistical error bar on $\se$ by a factor
of $0.5-0.7$ between $\L=0.2\iMpch$ and $\L=0.25\iMpch$ however (the statistical
error is on the order of 0.1\% or less for the latter cutoff, and not
visible in the plots).

Having improved substantially on the previous results of \cite{paperIIb}, this is likely to be the most precise inference (in terms of statistical as well as systematic error) of cosmological parameters \emph{from purely nonlinear information} in biased tracers of large-scale structure---albeit fixing the phases and other cosmological parameters.

The connection to observations will require the incorporation of redshift-space distortions and the window function. We have argued here that the latter is straightforward within the real-space formulation. We plan to investigate this next. Regarding the former, this is likewise straightforward within the EFT likelihood framework, by transforming the deterministic field $\dgdet$ to redshift space, and incorporating the Jacobian \cite{Cabass:2020jqo}. The Lagrangian forward model employed here is in fact ideally suited for this task. We leave an implementation of this to future work as well.

\acknowledgments

I would like to thank Tobias Baldauf, Giovanni Cabass, Oliver Hahn, Donghui Jeong, Elisabeth Krause, and Marcel Schmittfull for discussions, and Giovanni Cabass, Dragan Huterer, and Elisabeth Krause for comments that helped improve the draft significantly. 
I acknowledge support from the Starting Grant (ERC-2015-STG 678652) ``GrInflaGal'' of the European Research Council.

\clearpage

\appendix

\section{Supplementary figures}
\label{app:figs}

\begin{figure*}[thbp]
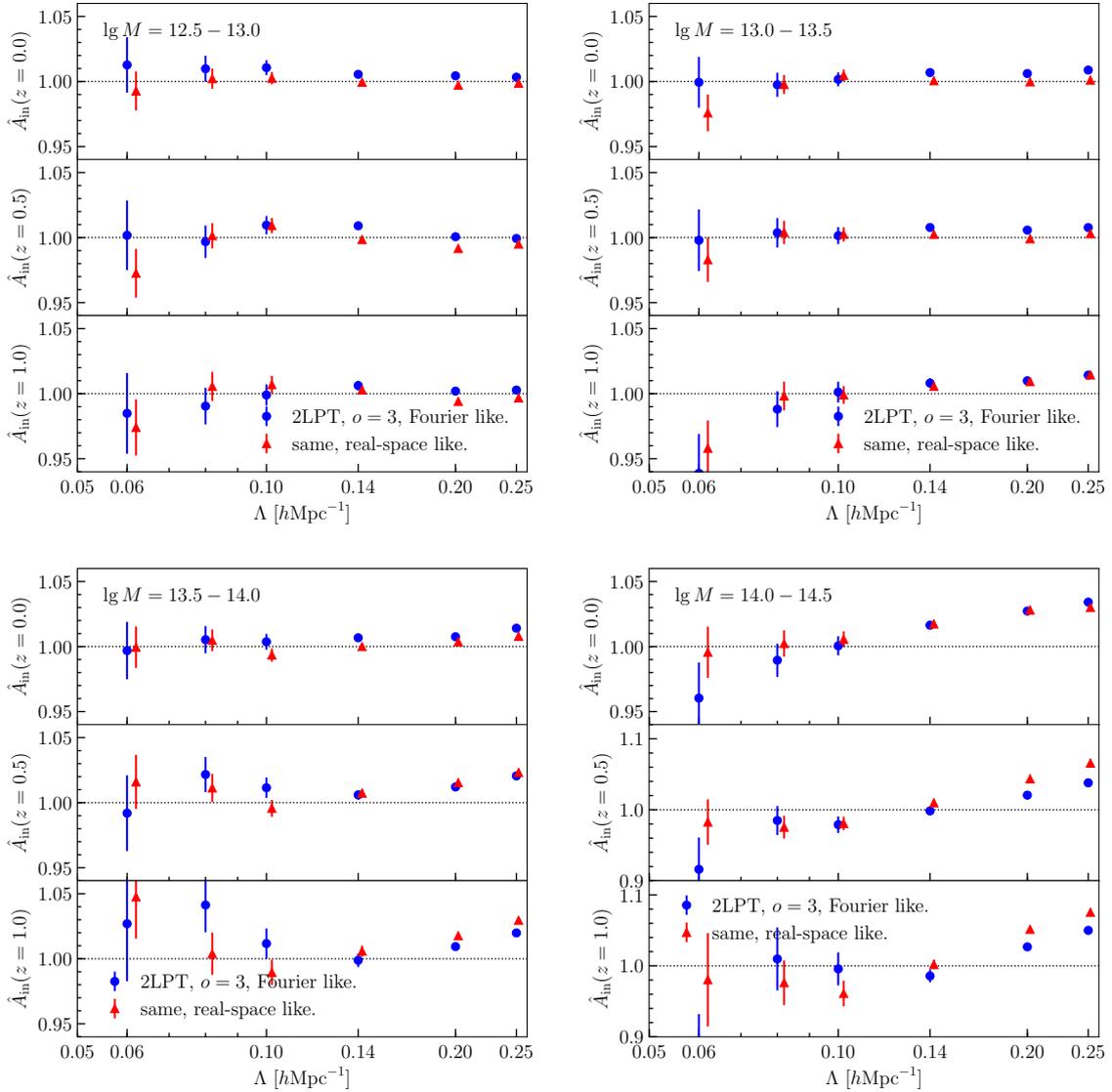
%
  \fourmassbinplots{sigma8_mlemargFourier_2LPT3D_prod}
  \cprotect\caption{
    Same as \reffig{LPTprod_3d}, but comparing results from real-space and Fourier-space likelihoods. Here, a 2LPT forward model using a $o=3$ bias expansion is used, corresponding to the third-order case of \cite{paperIIb} (but using a Lagrangian rather than Eulerian bias expansion). Results here and in all following figures are for run 1.
    \label{fig:LPTprod_2dF}}
\end{figure*}%

\begin{figure*}[thbp]
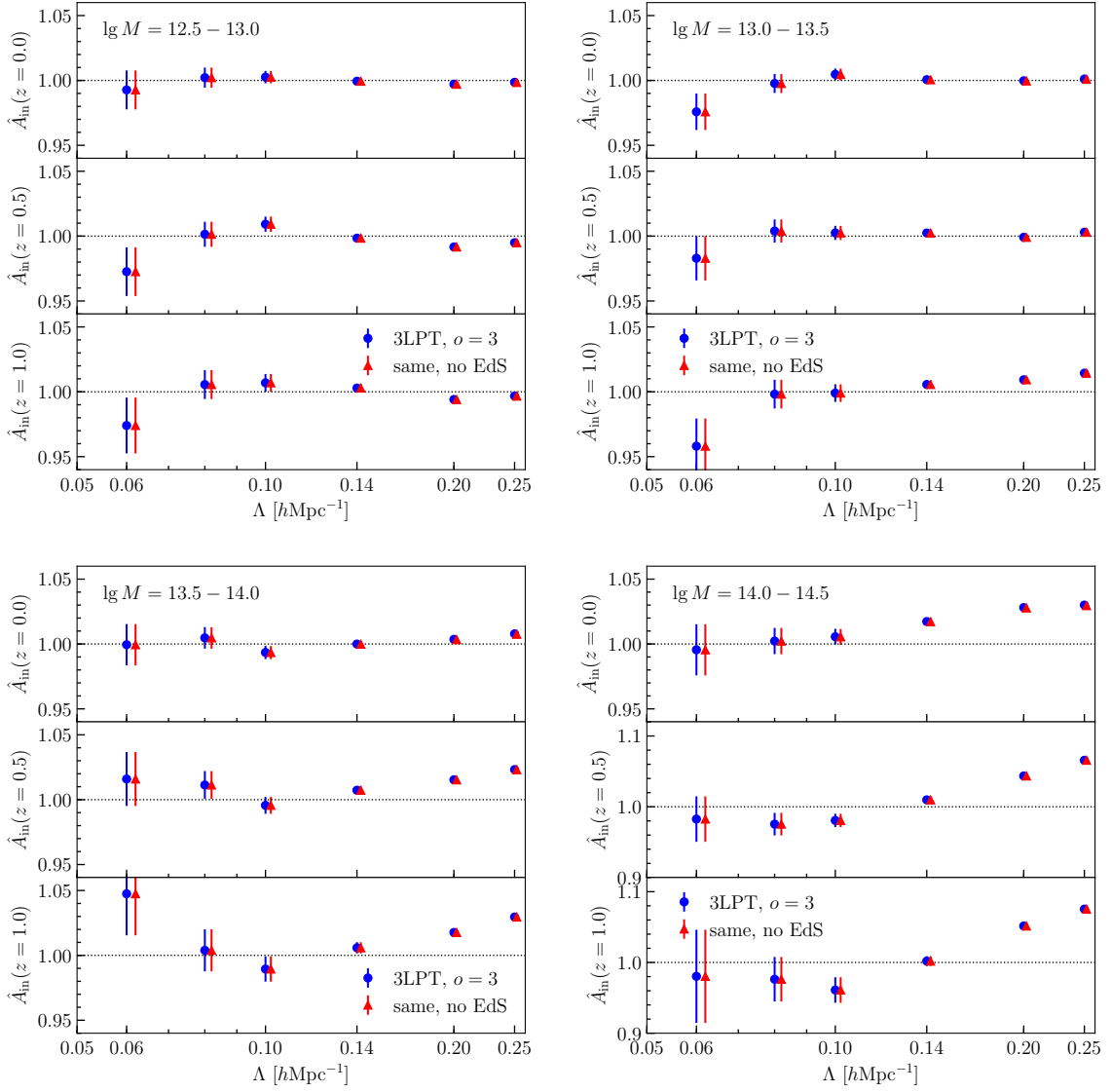
%
  \fourmassbinplots{sigma8_mlemargRealConstBEdS_3LPT3D_prod}
  \cprotect\caption{Effect of relaxing the assumption of the EdS approximation in the LPT forward model. The results are essentially unchanged.
    \label{fig:LPTprod_3d_BEdS}}
\end{figure*}%

\begin{figure*}[htbp]
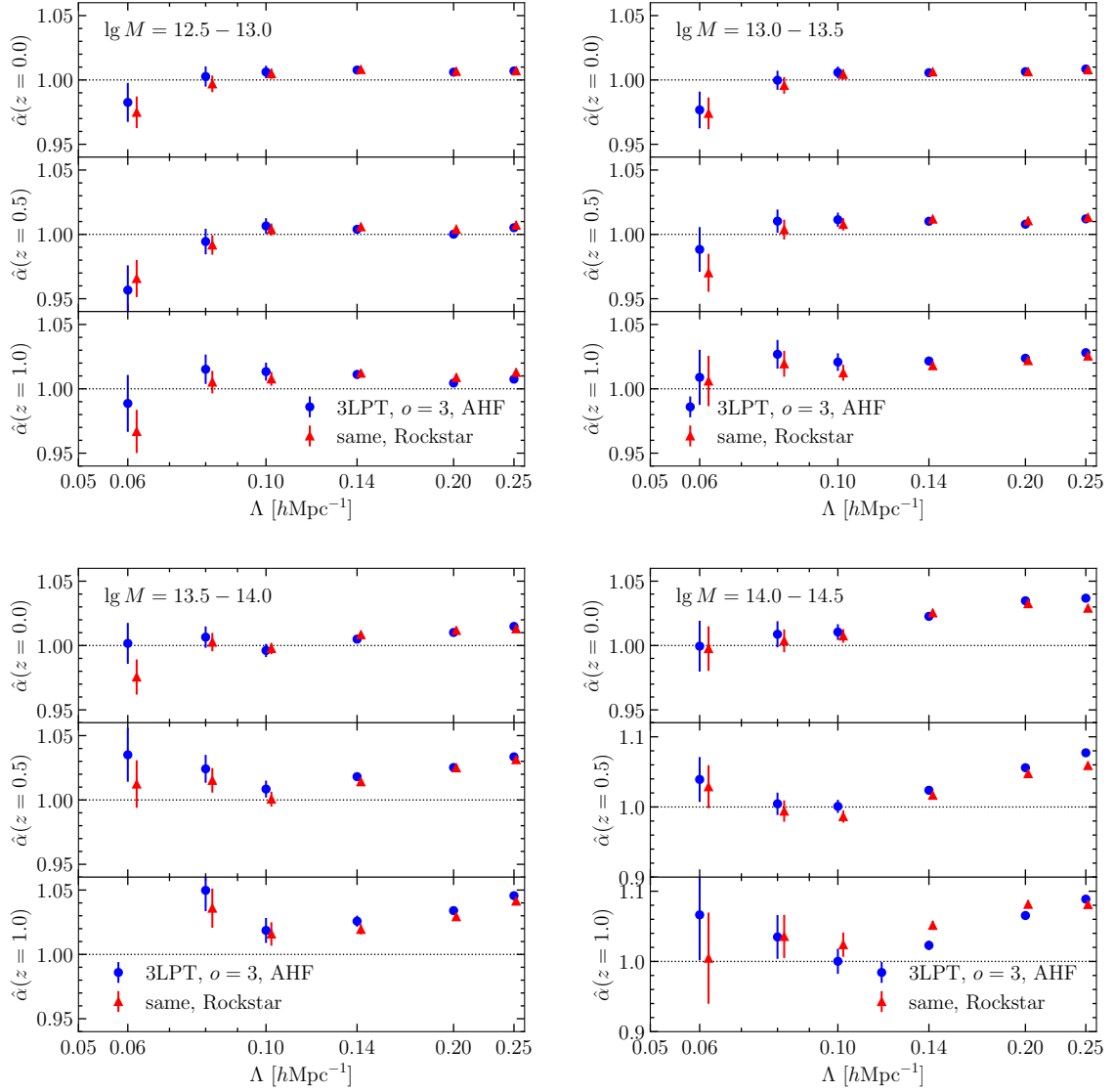
%
  \fourmassbinplots{sigma8_mlemargRealConst_3LPT3D_rockstar}
  \cprotect\caption{Results for the same likelihood as in \reffig{LPTprod_3d}, comparing the results of the fiducial \textsc{AHF} halo catalog with \textsc{Rockstar} halos. Note that both halo samples shown here were identified in simulations with $z_{\rm in}=99$. 
    \label{fig:rockstar}}
\end{figure*}%

\clearpage

\section{Grid reduction in Fourier space}
\label{app:reducesharpk}

An essential step in the real-space likelihood computation is the reduction
from a Fourier-space grid with $k_{\rm Ny} \gg \L$ to one where $k_{\rm Ny}=\L$.
We now describe how this is implemented. First, we determine the size of the
reduced grid via
\be
N_g^{\rm red} = 2 \left\lfloor\frac{\L L_{\rm box}}{2\pi} \right\rfloor\,,
\label{eq:Ngred}
\ee
where the floor function reduces $N_g^{\rm red}$ to the next even number,
which is desirable for numerical reasons. Thus, the actual cutoff is
slightly smaller than $\L$; in practice this difference is at most
$0.006 \iMpch$ for the box size considered here.

For all Fourier modes $\vk$ with $k_i < k_{\rm Ny}^{\rm red}/2$, where
$k_i$ denote the Cartesian components of $\vk$, we then
simply copy over the Fourier modes from the larger grid, scaling them
by the discrete FFT normalization factor $(N_g^{\rm red}/N_g)^3$.
For these modes the Hermitianity of the resulting field ($\d(-\vk) = \d^*(\vk)$
for a field $\d$) is ensured by way
of the Hermitianity of the input field on the full grid.
The modes on the Nyquist planes, for which $k_i = \pm k_{\rm Ny}^{\rm red}/2$
for at least one $i = 1,2,3$,
need to be handled slightly differently, because several (nonzero) modes of
the full grid are mapped onto the same mode on the reduced grid.

First, we set the imaginary
part of these modes to zero, as required for Hermitianity.
Second,
we multiply the real part of each mode on the Nyquist planes by a factor
$\sqrt{w}$, where $w$ counts the number of modes on the full grid that
are mapped onto the given mode on the reduced grid. Specifically,
\begin{itemize}
\item $w=2$ for modes on the faces of the Nyquist cube (a single component $k_i = \pm k_{\rm Ny}^{\rm red}/2$), since three of the six faces are represented on the reduced grid;
\item $w=4$ for modes on the edges (two components $k_i, k_j = \pm k_{\rm Ny}^{\rm red}/2$ with $i\neq j$), since 3 out of 12 edges are represented on the reduced grid;
\item $w=8$ for the single reduced-grid mode on the corner (all three components equal to $\pm k_{\rm Ny}^{\rm red}/2$; at the center of the reduced grid in terms of memory layout), since all 8 corners are mapped onto this mode.
\end{itemize}

This procedure ensures that, in the ensemble average, the modes on the
reduced grid have the same power as that on the full grid, which they
should since all modes that are nonzero on the full grid should be represented
on the reduced grid. 
We have verified that for a field $\d(\vk)$, the norm $\sum_{\vk} |\d(\vk)|^2$
agrees to a fractional precision of order $1/N_g^{\rm red}$ between the full and reduced grids, the
residual deviation being random fluctuations due to the modes on the
Nyquist planes. 

Finally, we ensure the Hermitianity of the resulting field by setting
$\d(-\vk) = \d^*(\vk)$. We do this by running only over the half-volume
with $k_x \geq 0$, where the modes with $k_x < 0$ are determined by the
Hermitianity condition. This procedure also allows for efficient
parallelization by dividing up the $k_x$ loop over threads (where the
otherwise problematic $k_x=0$ plane is assigned to a single thread).

Finally, the effective number of modes on the reduced grid is
\be
N_{\rm modes} = (N_g^{\rm red})^3 + 3 (N_g^{\rm red})^2 + 3 N_g^{\rm red} + 1.
\ee

\section{Halo catalogs}
\label{app:halos}

All numerical tests presented here are based on a set of
N-body simulations analagous to those used in \cite{paperII,paperIIb}, which were presented in
\cite{2017MNRAS.468.3277B}. They are generated using
\textsc{GADGET-2} \cite{2005MNRAS.364.1105S} for a flat $\Lambda$CDM cosmology
with parameters $\Omega_\mathrm{m} = 0.3$, $n_\mathrm{s} = 0.967$, $h = 0.7$,
and $\se = 0.85$, a box size of $L = 2000 \, h^{-1}
\mathrm{Mpc}$, and $1536^3$ dark matter particles of mass
$M_\mathrm{part} = 1.8 \times 10^{11} \, h^{-1} M_{\odot}$.
Two realizations are available, which we refer to as ``run 1'' and ``run 2.''
We do not use the simulations generated for \cite{2017MNRAS.468.3277B},
but instead have rerun them for the same initial phases but a later
starting redshift of $z_{\rm in}=24$ instead of 99 (see \refsec{Nbody}).

Dark matter halos
were subsequently identified at different redshifts as spherical
overdensities \cite{1974ApJ...187..425P, 1992ApJ...399..405W,1994MNRAS.271..676L} applying the Amiga Halo Finder algorithm (\textsc{AHF})
\cite{2004MNRAS.351..399G, 2009ApJS..182..608K} with an
overdensity threshold of $200$ times the background matter density.

For comparison, we have also generated halo catalogs using
\textsc{Rockstar} \cite{rockstar} (\refsec{rockstar}).
We used the same $\Delta=200b$ mass definition and the option
\verb!STRICT_SO_MASSES!.

\bibliographystyle{JHEP}
\bibliography{bibliography}

\end{document}